\documentclass[twocolumn,superscriptaddress,preprintnumbers,prb,citeautoscript]{revtex4}

\usepackage{graphicx}
\usepackage{amssymb}
\usepackage{amsfonts}
\usepackage{subfig}
\usepackage{dcolumn}
\usepackage{bm}
\usepackage{float}
\usepackage{footnote}
\usepackage{dcolumn}
\usepackage{amsmath}
\usepackage{amsfonts}
\usepackage{bm}
\usepackage{xcolor}
\usepackage{epsfig}
\usepackage{amssymb}
\usepackage{latexsym}
\usepackage[colorlinks,linkcolor=blue]{hyperref}
\usepackage{soul}
\usepackage{caption}
\newcommand{\be}{\begin{equation}}
\newcommand{\ee}{\end{equation}}
\newcommand{\bea}{\begin{eqnarray}}
\newcommand{\eea}{\end{eqnarray}}

\newcommand{\rd}{\mbox{d}}
\newcommand{\ri}{\mbox{i}}

\newcommand{\lrangle}[1]{\left< #1\right>}

\newcommand{\overbar}[1]{\mkern 1.5mu\overline{\mkern-1.5mu#1\mkern-1.5mu}\mkern 1.5mu}

\numberwithin{equation}{section}

\newcommand*{\citen}[1]{%
  \begingroup
    \romannumeral-`\x 
    \setcitestyle{numbers}%
    \cite{#1}%
  \endgroup   
}

\begin{document}

\title{Two types of superconducting pairs in stripe-ordered La$_{2-x}$Ba$_{x}$CuO$_4$ ($x=1/8$): evidence from the resistivity measurements}
\author{Tianhao Ren}
\email{tren@bnl.gov}
\affiliation{Condensed Matter Physics and Materials Science Division, Brookhaven National Laboratory, Upton, New York 11973, USA}
\author{Pedro M. Lozano}
\affiliation{Condensed Matter Physics and Materials Science Division, Brookhaven National Laboratory, Upton, New York 11973, USA}
\affiliation{Department of Physics and Astronomy, Stony Brook University, Stony Brook, New York 11794-3800, USA}
\author{Qiang Li}
\affiliation{Condensed Matter Physics and Materials Science Division, Brookhaven National Laboratory, Upton, New York 11973, USA}
\affiliation{Department of Physics and Astronomy, Stony Brook University, Stony Brook, New York 11794-3800, USA}
\author{Genda Gu}
\affiliation{Condensed Matter Physics and Materials Science Division, Brookhaven National Laboratory, Upton, New York 11973, USA}
\author{Alexei M. Tsvelik}
\affiliation{Condensed Matter Physics and Materials Science Division, Brookhaven National Laboratory, Upton, New York 11973, USA}
\begin{abstract}
Recent angle-resolved $c$-axis resistivity measurements of the stripe-ordered La$_{2-x}$Ba$_x$CuO$_4$ (LBCO) with $x=1/8$ revealed an unexpected dependence on the direction of the in-plane magnetic field. We argue that these and other available data for the $c$-axis transport point to the existence of superconducting pairs of two different types in the $x=1/8$ LBCO below the stripe ordering temperature. The pairs of one type  carry finite momentum and are confined to the Cu-O planes; the pairs of other type (probably the conventional $d$-wave with zero momentum) propagate along  narrow conducting channels traversing the sample in the $c$-axis direction. The evidence for this comes  from the  observed exponential temperature dependence of the $c$-axis resistivity $\rho_c(T)$ which we attribute to the thermally excited slips of the superconducting phase and flux flows. We present a simple theory to fit the observed  $\pi/2$-periodic dependence of $\rho_c$ on the direction of  the in-plane magnetic field and the other data.
\end{abstract}

\maketitle

\setstcolor{red}

\section{Introduction}
``It is a riddle, wrapped in a mystery, inside an enigma; but perhaps there is a key.'' These words of Winston Churchill he uttered in an entirely different context can be used to characterize the situation with the cuprates where enigma indeed contains in itself many layers. These structurally simple materials display a breathtaking complexity of properties, ranging from two-dimensional superconductivity \cite{TrTs,Tr,theory} to magnetically driven anomalous resistive states \cite{tranquada,dragana1,dragana2}. In this paper we discuss the stripe-ordered La$_{2-x}$Ba$_x$CuO$_4$ (LBCO) with $x=1/8$, with additional insights gained from recent angle-resolved $c$-axis resistivity measurements in the presence of an in-plane magnetic field \cite{PhysRevB.106.174510}.
  
\begin{figure}[htp!]
	\centering
	\includegraphics[width=0.8\linewidth]{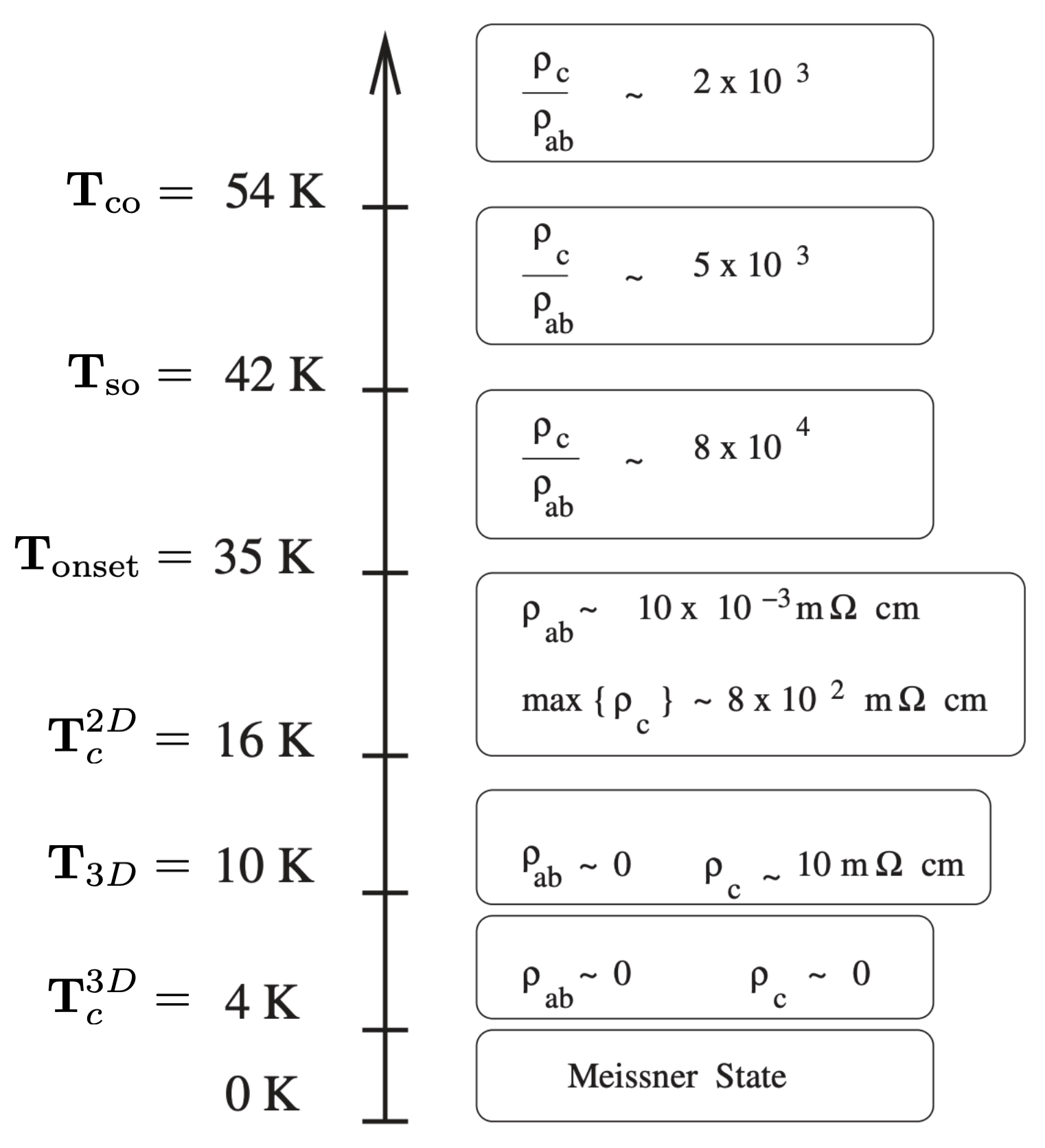}
	\caption{The relation between different temperature scales in the $x=1/8$ doped La$_{2-x}$Ba$_x$CuO$_4$ (LBCO), together with the transport regimes. Figure is adapted from Ref. \protect\citen{theory}.}
	\label{fig:temp}
\end{figure}

The most remarkable property of the stripe-ordered LBCO, discovered back in 2007, is the two-dimensional superconductivity \cite{TrTs,Tr}. The experimental measurements were  performed on the charge ordered low temperature tetragonal (LTT) phase in 1/8 doped LBCO. This phase has a crystal structure with two Cu-O layers per unit cell at $T \leq T_{\text{co}}$ = 54 K, where $T_{\text{co}}$ is the charge ordering temperature. In alternative layers, the charge stripes run along the $x$ or $y$ axis orthogonally. In the next-nearest neighboring layer, the parallel stripes are shifted by a half period to minimize the Coulomb interaction, further doubling the number of layers per unit cell. At temperatures below the spin ordering temperature $T_{\text{so}}$ = 42 K, the spins between each charge stripe have antiferromagnetic order along the stripe direction. The charge stripes act as anti-phase boundaries for neighboring spin stripes, further doubling the unit cell size in the direction perpendicular to the stripes. Below $T_{\text{onset}} < T_{\text{so}}$ the in-plane resistivity experiences a sharp drop and the system experiences  a crossover into the regime of strong 2D superconducting fluctuations  leading eventually to Berezinskii-Kosterlitz-Thouless (BKT) transition at $T_{c}^{2D} \sim 16$ K. Existence of such a sharp BKT transition is consistent with the theoretical expectations for layered superconductors without interlayer Josephson coupling \cite{RevModPhys.66.1125}.  In mean field theories $T_{\text{onset}}$ would be a transition temperature, but in reality it marks the entrance to the regime of phase incoherence dominated by thermally excited vortices. Also, $T_{\text{so}}$ does not depend on magnetic field, but $T_{\text{onset}}$ does, which strongly suggests the existence of preformed pairs presumably located on the stripes, protected by the spin gap \cite{theory,Soto_2015,Sopik_2015}. \normalsize The $c$-axis resistivity vanishes below $T_{3D}\sim 10$ K, while the 3D superconductivity, characterized by the Meissner state, is achieved only below $T_c^{3D} \sim 4$ K. The relation between different temperature scales are shown in Fig. \ref{fig:temp}.
 
The theoretical explanation suggested soon after the original discovery \cite{theory} portrayed the superconducting state as a quasi-condensate of pairs with a finite momentum, termed as pair density wave (PDW). Then its two dimensional character originates from the fact that momenta of the pairs in adjacent layers have different orientations which frustrates the interlayer Josephson coupling. This conclusion is supported by the optical measurements which established the decoupling of the superconducting orders in adjacent copper oxide layers \cite{Agterberg}. The theory describes the stripe-ordered state of the doped antiferromagnet as an array of self-assembled rivers of charges separated by insulating regions. Each charge river is represented as a Luther-Emery liquid where the formation of the spin gap leads to the superconducting pairing; the pairing fields from different rivers are weakly coupled by the Josephson pair tunneling. To explain the finite momentum of the pairs, the theory in \cite{theory} posits the negative sign of the Josephson matrix element - an assumption for which so far little theoretical justification has been provided.

It was later found in \cite{tranquada,dragana1,dragana2} that when the magnetic field destroys superconductivity, a very peculiar resistive metallic state emerges which persists down to lowest temperatures and inherits such prominent feature of the superconductor as zero Hall response. The critical magnetic field is relatively small; the electrical resistance gradually increases with the magnetic field, and at field strength around $25-30$ T, the sheet resistance reaches a plateau at $R_{\square} \approx 2\pi \hbar/2e^2$. This is dubbed ``ultra-quantum metal", which lies between $T_{\text{onset}}$ and $T_{\text{so}}$, and the Josephson coupling between the stripes is suppressed and the superconducting fluctuations are one dimensional. There is a dichotomy in the theoretical explanation of this state between the quasiparticle picture \cite{tsvelik16,tsvelik19,lee} and the Bose metal picture \cite{PhysRevB.60.1261,Phillips_2018,RenTs}, where the transport is carried by quasiparticles in the former and by the incoherent pairs in the latter.

The previous experimental and theoretical efforts have concentrated mostly on understanding of physics of the $ab$ planes, while the $c$-axis transport and the possible mechanism for the Meissner state have not attracted much attention. This situation is changed by recent angle-resolved $c$-axis resistivity measurements \cite{PhysRevB.106.174510}, which attract attention to the puzzles related to  the dramatic difference between the $c$-axis transport and the $ab$-plane transport. The experimental findings strongly suggest the existence of two subsystems in the stripe-ordered LBCO, one being the 2D superconductivity in the form of PDW within the $ab$ planes, the other being the 1D superconductivity traversing along the $c$ axis. This is what we will describe in detail in the following sections.

Before proceeding, we want to make a clarification of the relation between the current paper and Ref. \citen{PhysRevB.106.174510} by the same group of authors. On the experimental side, the data shown later in Fig. \ref{fig:rhoab}, \ref{fig:rhoc}, and \ref{fig:lnrhoc} are taken from Ref. \citen{PhysRevB.106.174510}, and analyzed further in the current paper. The data shown later in Fig. \ref{fig:resis0}, \ref{fig:resis1}, \ref{fig:field}, and \ref{fig:field2} are exclusively reported in the current paper, although similar data as that of Fig. \ref{fig:resis0} and \ref{fig:resis1} for different temperatures and magnetic fields are already reported in Ref. \citen{PhysRevB.106.174510}. On the theoretical side, the existence of the two types of superconducting pairs based on the observed angle-dependent magneto-resistivity is briefly mentioned in Ref. \citen{PhysRevB.106.174510} with a promise to  discuss it  in detail in subsequent publications. The current paper serves as a follow-up to Ref. \citen{PhysRevB.106.174510} fulfilling this promise, providing theoretical details and discussing further important implications.

\section{Experimental Observations}

\begin{figure}[htp!]
	\centering
	\includegraphics[width=0.9\linewidth]{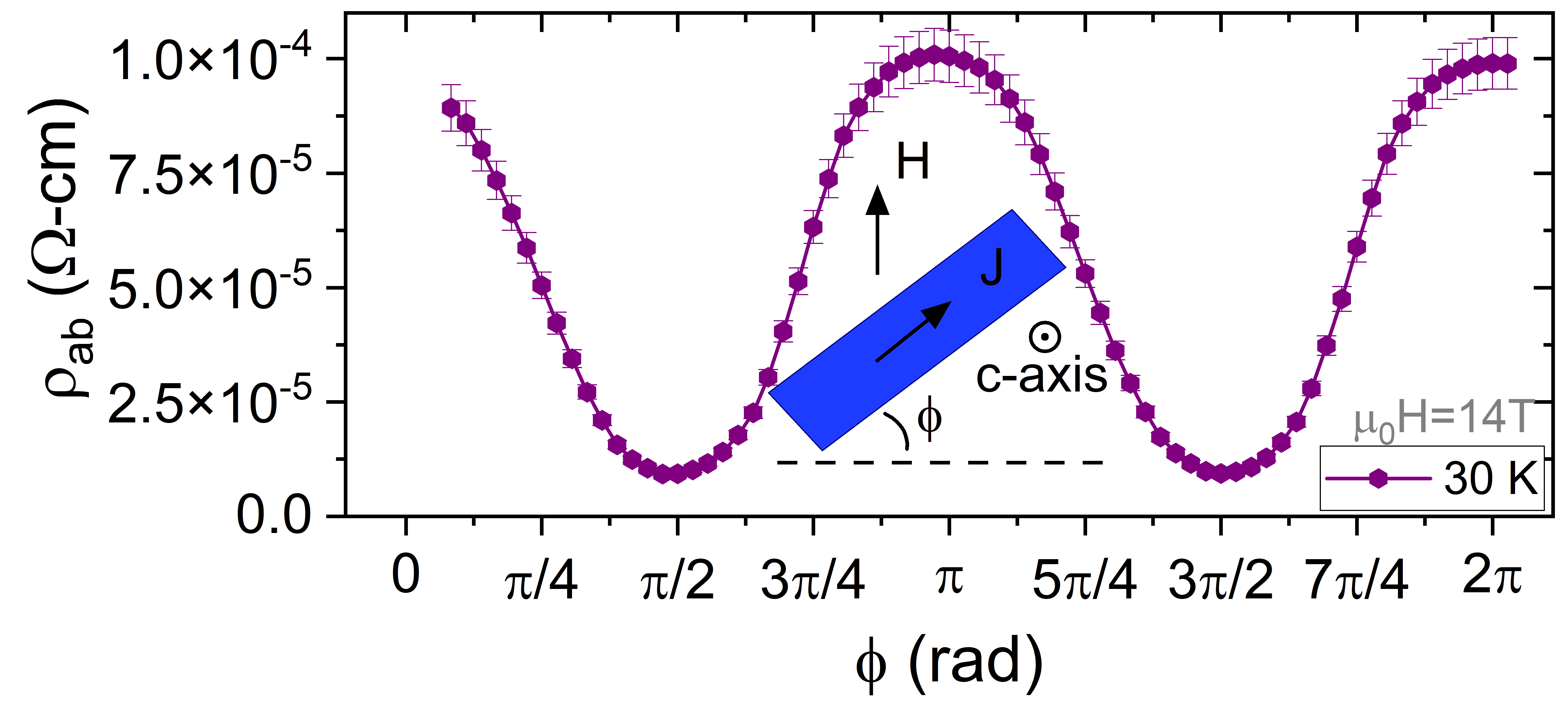}
	\caption{Characteristic measurements of the angle dependent in-plane resistivity $\rho_{ab}$ at $B=14$T, $T=30$K. $\rho_{ab}$ depends only on the angle between the in-plane magnetic field and the current direction, but not on the stripe directions. The angle dependence has a period of $\pi$. These data are originally reported in Ref. \protect\citen{PhysRevB.106.174510}.}
	\label{fig:rhoab}
\end{figure}

The experimental measurements are performed on the stripe-ordered LBCO, where the in-plane resistivity $\rho_{ab}$ and the $c$-axis resistivity $\rho_c$ are measured in presence of an in-plane magnetic field. The measurements are performed at different temperatures and different strengths of the magnetic field, with the direction of the magnetic field varied within the $ab$ plane. Before moving on to the behavior of the $c$-axis resistivity, we briefly address the behavior of the in-plane resistivity $\rho_{ab}$. Due to the layer decoupling, we can focus on a single layer for the leading-order behavior of $\rho_{ab}$. Below $T_{\text{onset}}$, the preformed pairs are protected by the spin gap. Above $T^{2D}_c$, there is a proliferation of vortices, which are thermally activated to produce a resistivity as \cite{Nelson,RevModPhys.59.1001}
\begin{equation}
\label{eq:rhoab}
	\rho_{ab} \propto \rho^n_{ab}\exp\left(-b\sqrt{\frac{T_{\text{onset}}-T}{T-T^{2D}_c}}\right),
\end{equation}
where $b$ is a dimensionless fitting parameter, and the normal resistivity $\rho^n_{ab}$ receives a contribution from the freed quasiparticles due to loss of pair coherence within the vortices. The temperature dependence of Eq. (\ref{eq:rhoab}) agrees qualitatively well with the experimental data \cite{theory}. The magnetic field dependence of Eq. (\ref{eq:rhoab}) hides itself in the normal resistivity $\rho^n_{ab}$:
\begin{equation}
\label{eq:rhoab2}
	\rho^n_{ab}=\frac{1+\omega^2_c\tau^2}{\sigma_D},
\end{equation}
where $\sigma_D$ is the Drude conductivity of the quasiparticles, $\tau$ is the elastic scattering relaxation time, and $\omega_c\propto B_{\perp}$ is the cyclotron frequency, with $B_{\perp}=B\cos\phi$ being the component of the in-plane magnetic field perpendicular to the current. The longitudinal component $B_{\parallel}=B\sin\phi$ exerts no force on the current, as can be seen from the Lorentz formula $\bm{F}_L=\bm{j}\times\bm{B}$. This renders $\rho_{ab}$ dependent on the angle between the in-plane magnetic field and the current (but not on the stripe directions), with a period of $\pi$. This is consistent with the experimental data, shown in Fig. \ref{fig:rhoab}. 

\begin{figure}[htp!]
	\centering
	\includegraphics[width=0.8\linewidth]{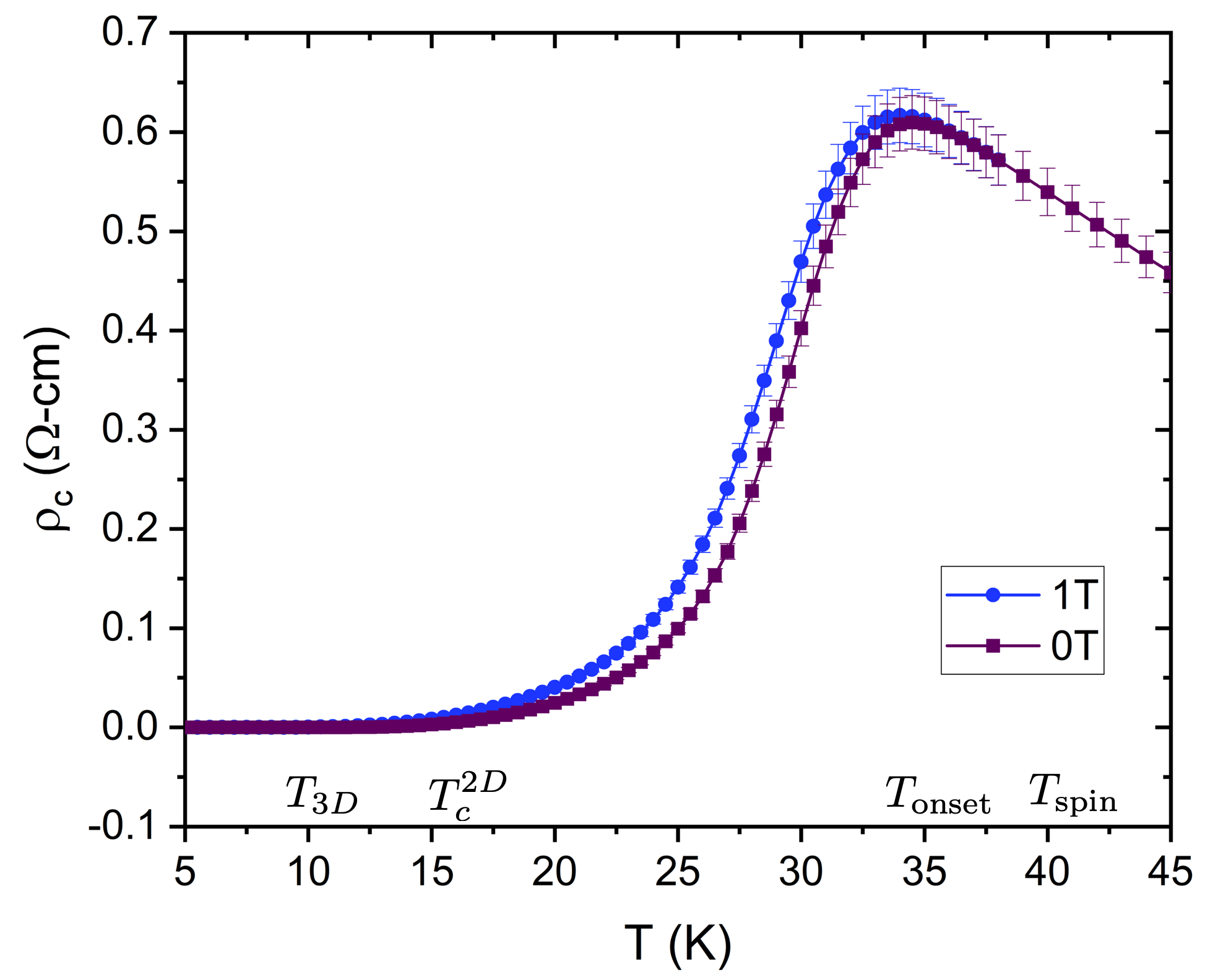}
	\caption{The temperature dependence of the $c$-axis resistivity $\rho_c$ at $B=0$T and $B=1$T, where the in-plane magnetic field is along the Cu-Cu bond. $\rho_c(T)$ displays a maximum around the stripe ordering temperature, and a thermally activated form below $T_{\text{onset}}$. The temperature scales are also shown, where $T^{3D}_c$ (below 5K) and $T_{\text{co}}$ (above 45K) are not shown. These data are originally reported in Ref. \protect\citen{PhysRevB.106.174510}.}
	\label{fig:rhoc}
\end{figure}

\begin{figure}[htp!]
	\centering
	\includegraphics[width=0.8\linewidth]{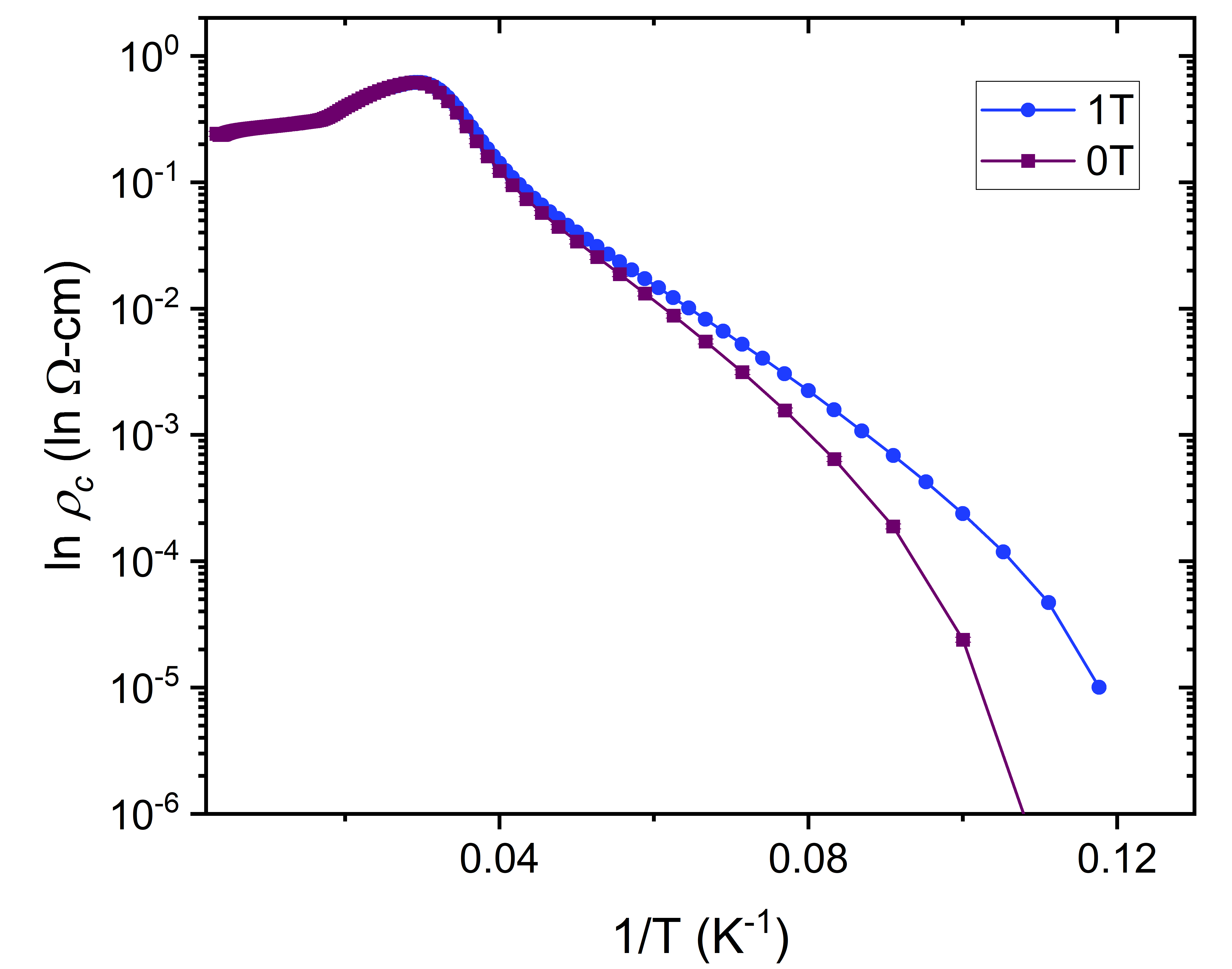}
	\caption{The temperature dependence of the $c$-axis resistivity at $B=0$T and $B=1$T (the in-plane magnetic field is along the Cu-Cu bond) shown on semi-logarithmic scale as $\ln\rho_c(T)$ vs $1/T$. The error bars are hidden inside the data points and can be seen if one enlarges the figure.  These data are originally reported in Ref. \protect\citen{PhysRevB.106.174510}.}
	\label{fig:lnrhoc}
\end{figure}

Now we proceed to the behavior of the $c$-axis resistivity. As mentioned above, in a zero magnetic field, the $c$-axis transport is resistive above $T_{3D}\sim 10$ K. As shown in Fig. \ref{fig:rhoc}, $\rho_c(T)$ displays a characteristic maximum around the stripe ordering temperature and a thermally activated form $\rho_c\sim \exp(-\Delta/T)$ below $T_{\text{onset}}$. This brings to mind the Ambegaokar-Halperin formula  \cite{PhysRevLett.22.1364,RevModPhys.38.298} describing the resistivity in thin superconducting wires, where the dissipation is generated by thermally activated phase slips:
\begin{equation}
\label{eq:rhoc1}
	\rho^{\text{slip}}_c \propto \rho^n_c\frac{\Omega}{T}\exp\left( -\Delta^{\text{slip}}/T \right), 
\end{equation}
where $\rho^n_c$ is the normal resistivity due to the interlayer quasiparticle hopping, $\Omega$ is a characteristic frequency, and the activation energy $\Delta^{\text{slip}}$ can be estimated as $\Delta\sim H_{c1}d^2\xi_c$ according to the Langer-Ambegaokar-McCumber-Halperin theory \cite{PhysRev.164.498,PhysRevB.1.1054}, provided the thickness $d$ of the wire is much smaller than the $c$-axis coherence length $\xi_c$, and $H_{c1}$ is the bulk lower critical magnetic field. The temperature dependence of Eq. (\ref{eq:rhoc1}) agrees qualitatively well with our experimental data shown in Fig. \ref{fig:rhoc} and \ref{fig:lnrhoc}, where the same data of Fig. \ref{fig:rhoc} is plotted on semi-logarithmic scale in Fig. \ref{fig:lnrhoc} for a better visualization of the exponential dependence in Eq. (\ref{eq:rhoc1}).

\begin{figure}
	\centering
	\includegraphics[width=.9\linewidth]{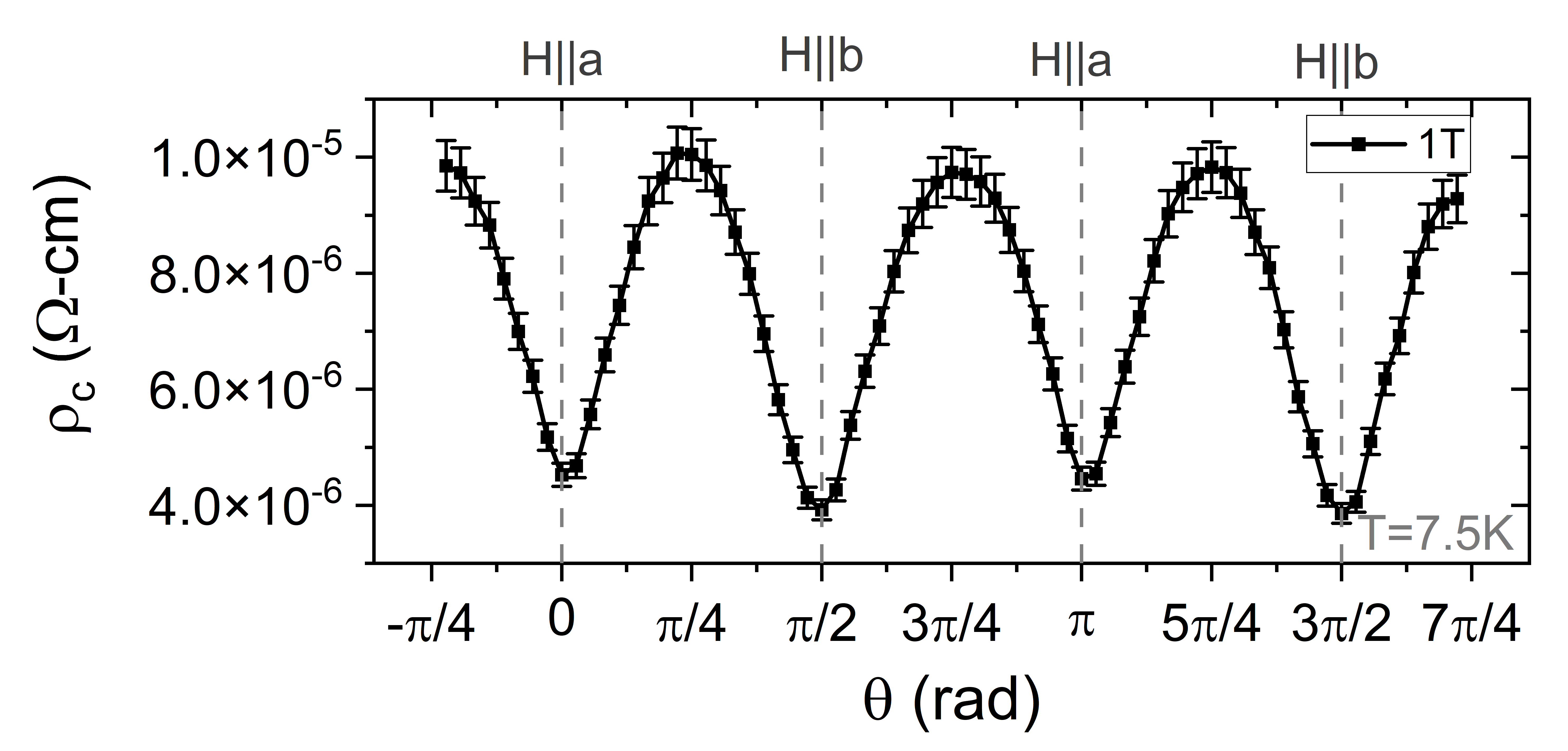}
	\caption{Characteristic measurements of the angle-resolved $c$-axis resistivity at $B=1$T, $T=7.5$K. $\rho_c$ depends on the angle between the in-plane magnetic field and the stripe direction. $a$ and $b$ are along the Cu-O-Cu bond, so the minima of $\rho_c$ appear where the magnetic field is along the directions of the stripes. The angle dependence has a period of $\pi/2$. Materials and methods for these original data are discussed in Appendix \ref{sec:MM}.}
	\label{fig:resis0}
\end{figure}

The recent experiments measured $\rho_c$ in the presence of an in-plane magnetic field \cite{PhysRevB.106.174510}. The measurements were performed both above and below $T_{\text{so}}$ as a function of the angle $\theta$ between an in-plane magnetic field and the stripe direction. Above $T_{\text{so}}$, $\rho_c$ shows no angular dependence. Below $T_{\text{so}}$,  $\rho_c$ exhibits an angular dependence with a period of $\pi/2$. The angular dependence does not experience any drastic change at $T_{\text{onset}}$. The $\rho_c$ minima appear at $0,\pi/2,\pi,3\pi/2$, where the magnetic field is along the directions of the stripes. The $\rho_c$ maxima appear at $\pi/4,3\pi/4, 5\pi/4, 7\pi/4$, where the magnetic field bisects the stripe directions in adjacent layers, or along the nodal directions of the $d$-wave order parameters. Characteristic experimental results are shown in Fig. \ref{fig:resis0}.

In the presence of  magnetic field, there is an extra contribution to $\rho_c$ due to the  thermally activated flux flows \cite{PhysRev.140.A1197,RevModPhys.66.1125}:
\begin{equation}
\label{eq:rhoc2}
	\rho^{\text{flow}}_c\propto \rho^n_c\frac{B}{H_{c2}}\exp\left(-\Delta^{\text{flow}}/T\right),
\end{equation}
where $B$ is the strength of the magnetic field, $H_{c2}$ is the bulk upper critical magnetic field, and the activation energy $\Delta^{\text{flow}}$ is generally smaller than $\Delta^{\text{slip}}$ (this can be seen from Fig. \ref{fig:lnrhoc}). Consequently, in the presence of magnetic field $\rho_c$  is dominated by $\rho^{\text{flow}}_c$. The temperature dependence of Eq. (\ref{eq:rhoc2}) agrees qualitatively well with the experimental data shown in Fig. \ref{fig:rhoc} and \ref{fig:lnrhoc}, while the angular dependence must be hidden inside the normal resistivity $\rho^n_c$ \footnote{Both Eq. (\ref{eq:rhoc1}) and (\ref{eq:rhoc2}) have the prefactor $\rho^n_c$, which encodes the dependence on the direction of the magnetic field. We assume that the bulk critical fields $H_{c1}$, $H_{c2}$ and the activation energies $\Delta^{\text{slip}}$, $\Delta^{\text{flow}}$ all have negligible dependence on the direction of the in-plane magnetic field.}. If the quasiparticles possess anisotropy compatible with the PDW configuration, the rotation of the PDW direction along the $c$ axis will come with a rotation of the axis of anisotropy, such that the quasiparticle hopping along the $c$ axis will produce the observed angular dependence of $\rho^n_c$, which will be shown later. This naturally explains the different angular dependences observed between $\rho_{ab}$ and $\rho_c$.

\section{The Phenomenological Model}

 The transport data described above indicate that above the 3D superconducting transition in the stripe-ordered LBCO there are two types of carriers responsible for the in-plane and $c$-axis transport respectively. These different pairs inhabit subspaces of different dimensions. One subspace, let us call it A, consists of two-dimensional arrays of superconducting stripes, where pairs with finite momentum condense. The superconducting pairs formed in the stripes of a given plane remain confined to this plane which determines the BKT character of the transition. The other subspace, call it B, includes areas where the pairing has a different symmetry (it is most likely a conventional $d$-wave one, which is common for cuprates \cite{Norman_2003} ) and hence the corresponding order parameter field does not couple to the one of the stripes. We assume that the subspace B consists of narrow channels oriented along the $c$ axis, probably formed  by columnar defects. These pairs are unable to achieve  phase coherence due to the phase slips and flux flows. Since due to the momentum mismatch there is no pair tunneling between A and B subspaces (otherwise it will break momentum conservation), the transport along the $c$-axis can remain resistive even below the BKT transition as seen experimentally. Since according to Eq. (\ref{eq:rhoc1}) and (\ref{eq:rhoc2}) the observed $\rho_c$ is proportional to the resistivity of the normal state, to explain the angular dependence of the magnetoresistivity, we have to study the normal state resistivity $\rho^n_c$. 

\begin{figure}[htp!]
	\centering
	\includegraphics[width=0.8\linewidth]{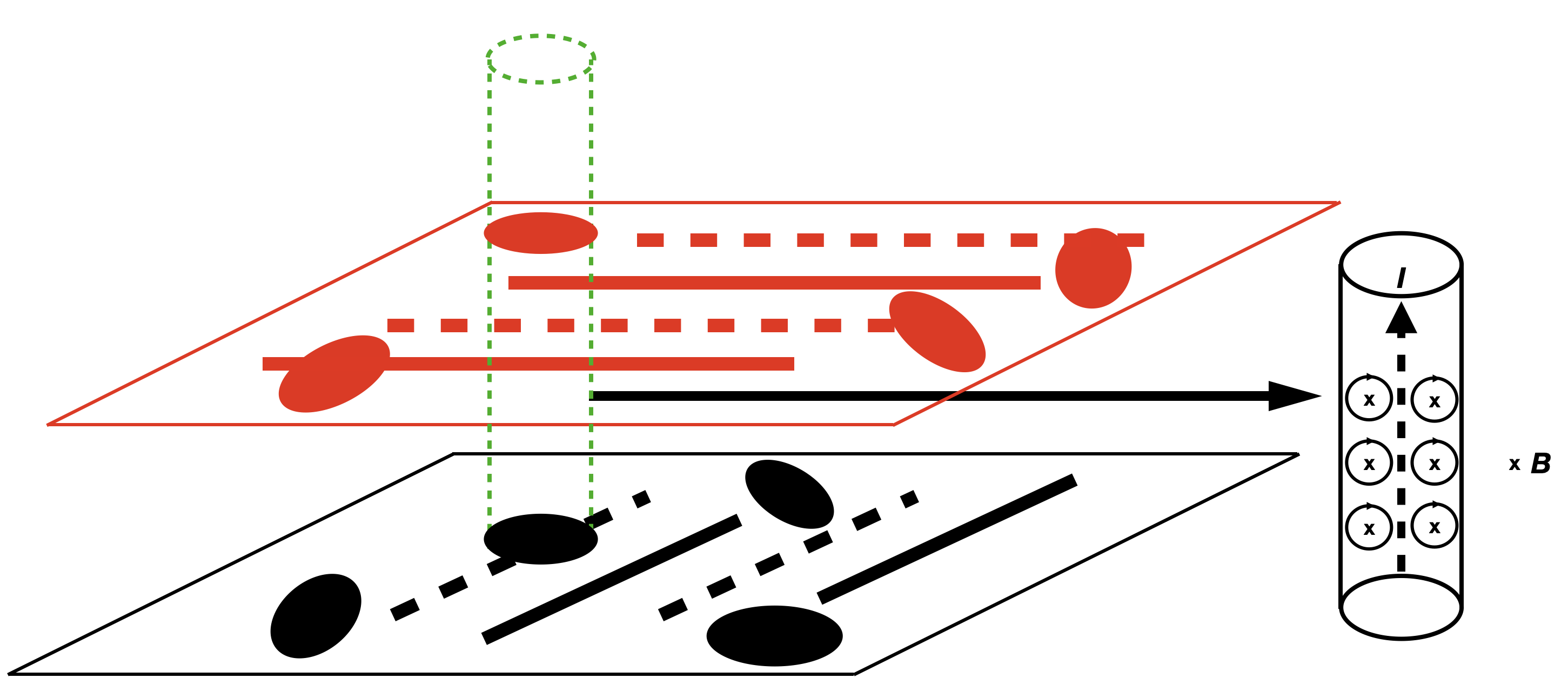}
	\caption{The two subsystems of the stripe-ordered LBCO, where within the $ab$ plane, we have the peculiar PDW configuration, and along the $c$ axis we have the superconducting wire. The superconducting wire is enlarged on the righthand side for illustration of the fluxes produced by the in-plane magnetic field.}
	\label{fig:thread}
\end{figure}

To estimate $\rho^n_c$ we adopt a simplified model of quasiparticle hopping to illustrate that the desired angular dependence of $\rho^n_c$ can be attributed to quasiparticles of the Dirac fermion type with anisotropy compatible with the peculiar PDW configuration shown in Fig. \ref{fig:thread}. At present we do not have any microscopic model to support this claim. However, we are forced to introduce the Dirac quasiparticles by the weight of the evidence coming from the experimental data. As demonstrated in Appendix \ref{sec:quadratic}, the scenario where quasiparticles has quadratic dispersion should be excluded. The discussion below provides further details.

\begin{figure}[htp!]
 	\centering
	\includegraphics[width=.9\linewidth]{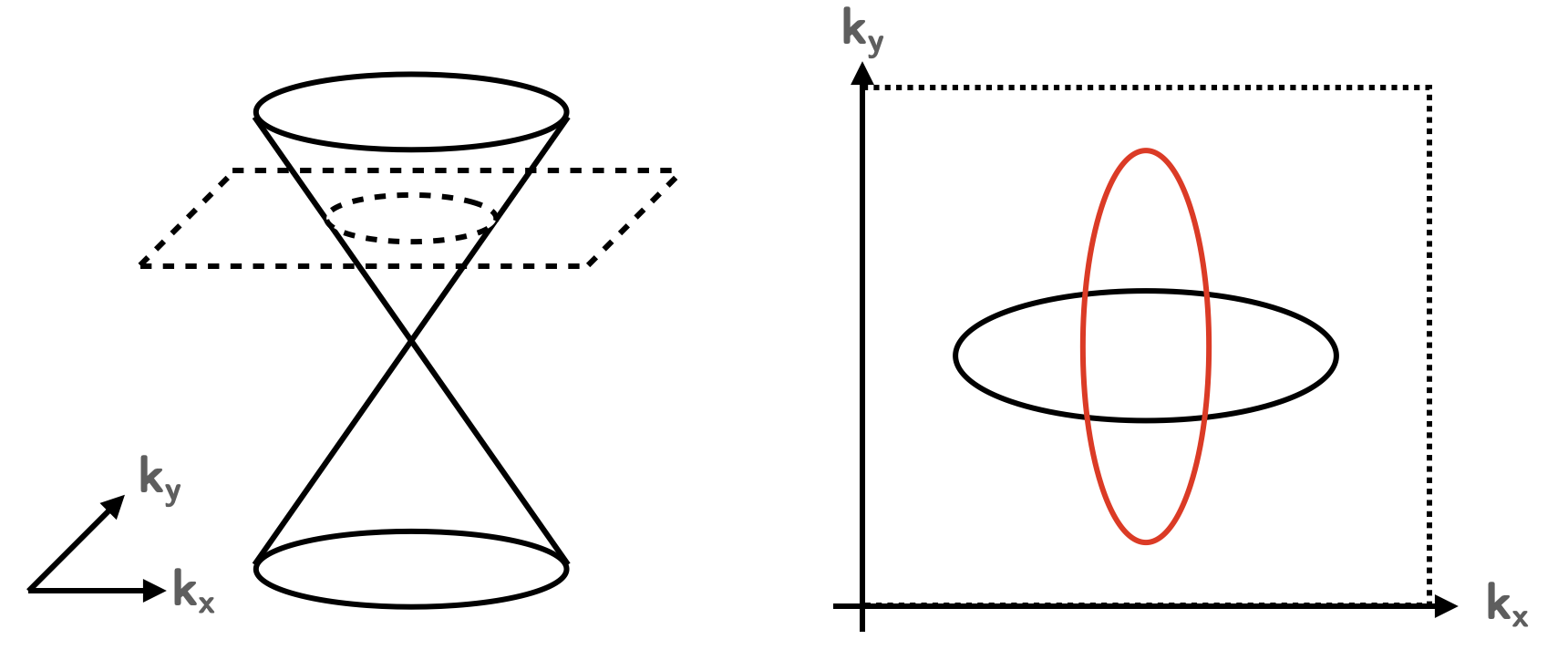}
	\caption{The quasiparticle of the Dirac fermion type with anisotropy compatible with the peculiar PDW configuration, where black and red ellipses correspond to the cross sections of the Dirac cones in nearest neighboring layers.}
	\label{fig:FS1}
\end{figure}

In the simplified model, we consider two layers in the unit cell instead of four \footnote{The parallel stripes in the next-nearest neighboring layer are shifted by a half period, doubling the number of layers per unit cell, but this does not have an effect on the rotation of the axis of anisotropy, and it is the stripe direction that matters for the oscillation of the magneto-resistivity, as shown by the experimental data.}. The simplest arrangement corresponds to Dirac fermions with anisotropy compatible with the peculiar PDW configuration as shown in Fig. \ref{fig:FS1}, where the main axes of the cross sections are along the directions of the stripes and rotated by 90 degrees in the nearest neighboring layer. Although the $d$-wave superconducting regions do not have the stripe order, they do still have the LTT structure, where there is an anisotropy of the Cu-O hopping.  In one direction, the Cu-O-Cu bonds are in line, while at 90 degrees there is a bend of $2\phi$ about 7 degrees (for an estimation of the tilt angle, refer to Fig. 15(a) of [\citen{PhysRevB.83.104506}]). The bend direction alternates from one O site to the next, and rotates 90 degrees between adjacent layers. This could be a potential mechanism responsible for the anisotropy investigated here.

\begin{figure}[htp!]
	\centering
	\includegraphics[width=.9\linewidth]{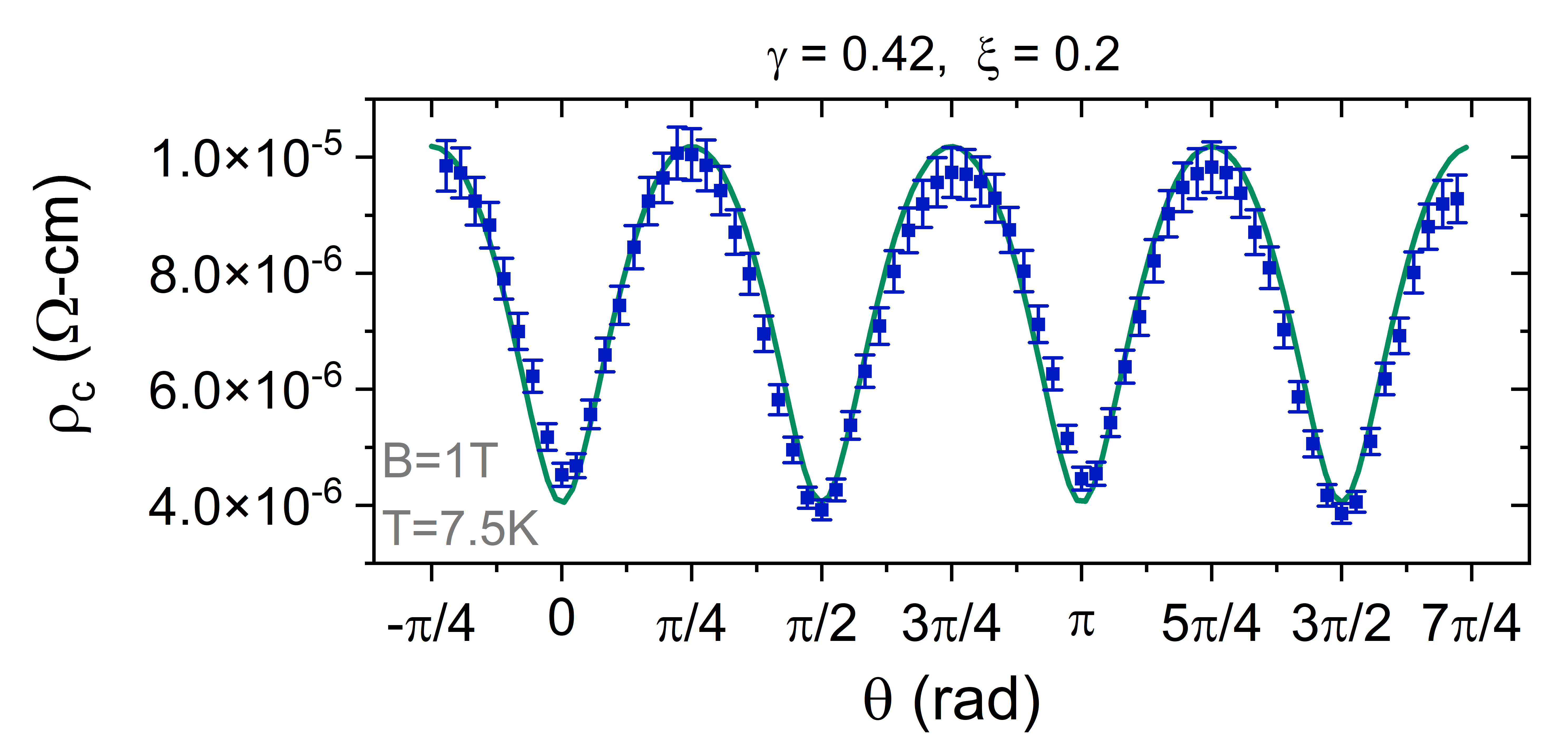}
	\caption{The fitting of the $c$-axis resistivity according to Eq. (\ref{eq:rhoc}), in correspondence to Fig. \ref{fig:resis0}. The blue solid line represents the theoretical prediction, while the orange solid dots represent the experimental data. Materials and methods for these original data are discussed in Appendix \ref{sec:MM}.}
	\label{fig:resis1}
\end{figure}

\begin{figure}[htp!]
	\centering
	\includegraphics[width=0.8\linewidth]{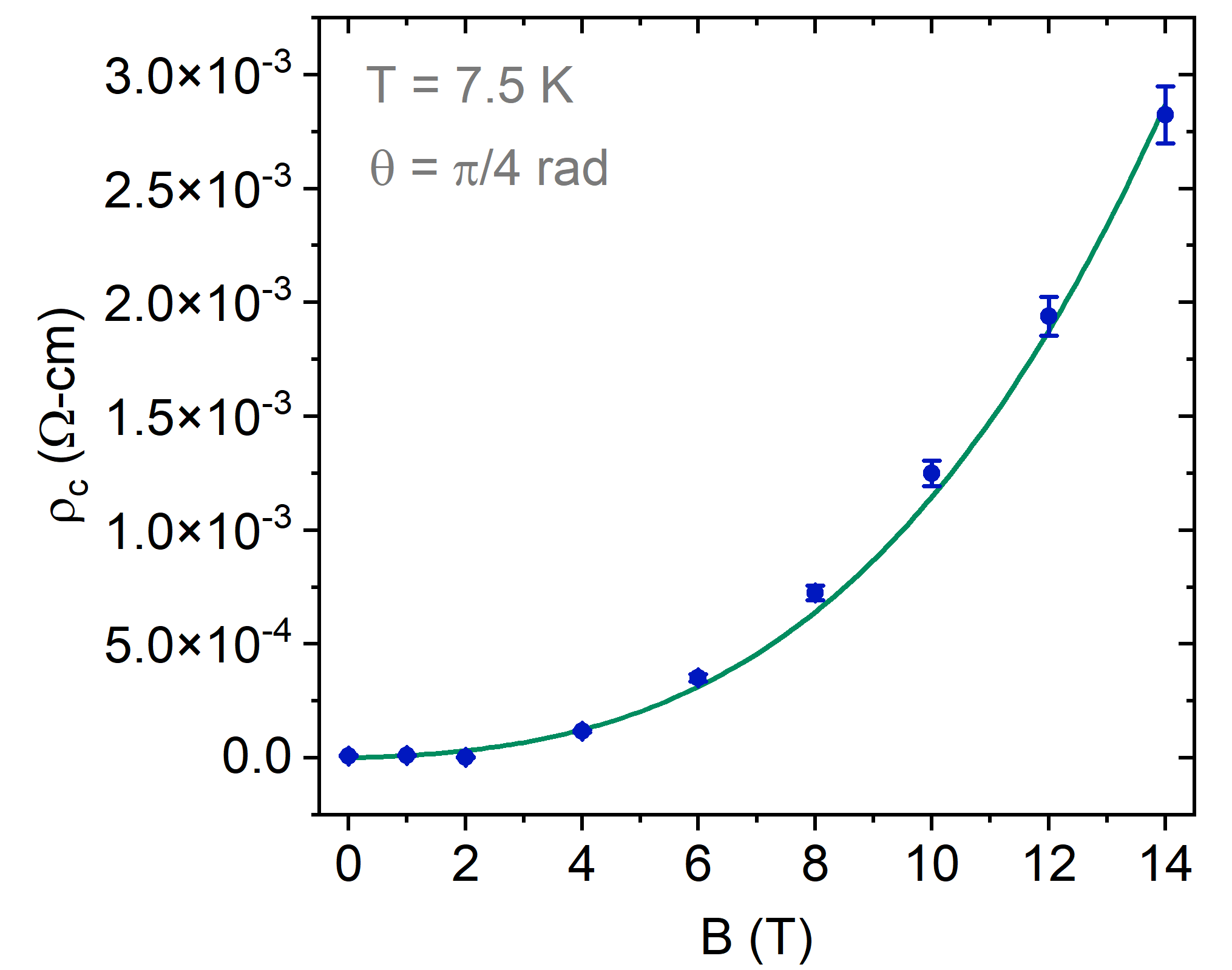}
	\caption{The dependence of the $c$-axis resistivity on the magnitude of the in-plane magnetic field, shown for a fixed direction of the magnetic field (in fact, the direction of resistivity maximum). The blue solid line is the theoretical prediction from Eq. (\ref{eq:rhoc}) and (\ref{eq:rhoc2}), and the orange solid dots are from the experimental data. Materials and methods for these original data are discussed in Appendix \ref{sec:MM}.}
	\label{fig:field}
\end{figure}

\begin{figure}[htp!]
	\centering
	\includegraphics[width=0.8\linewidth]{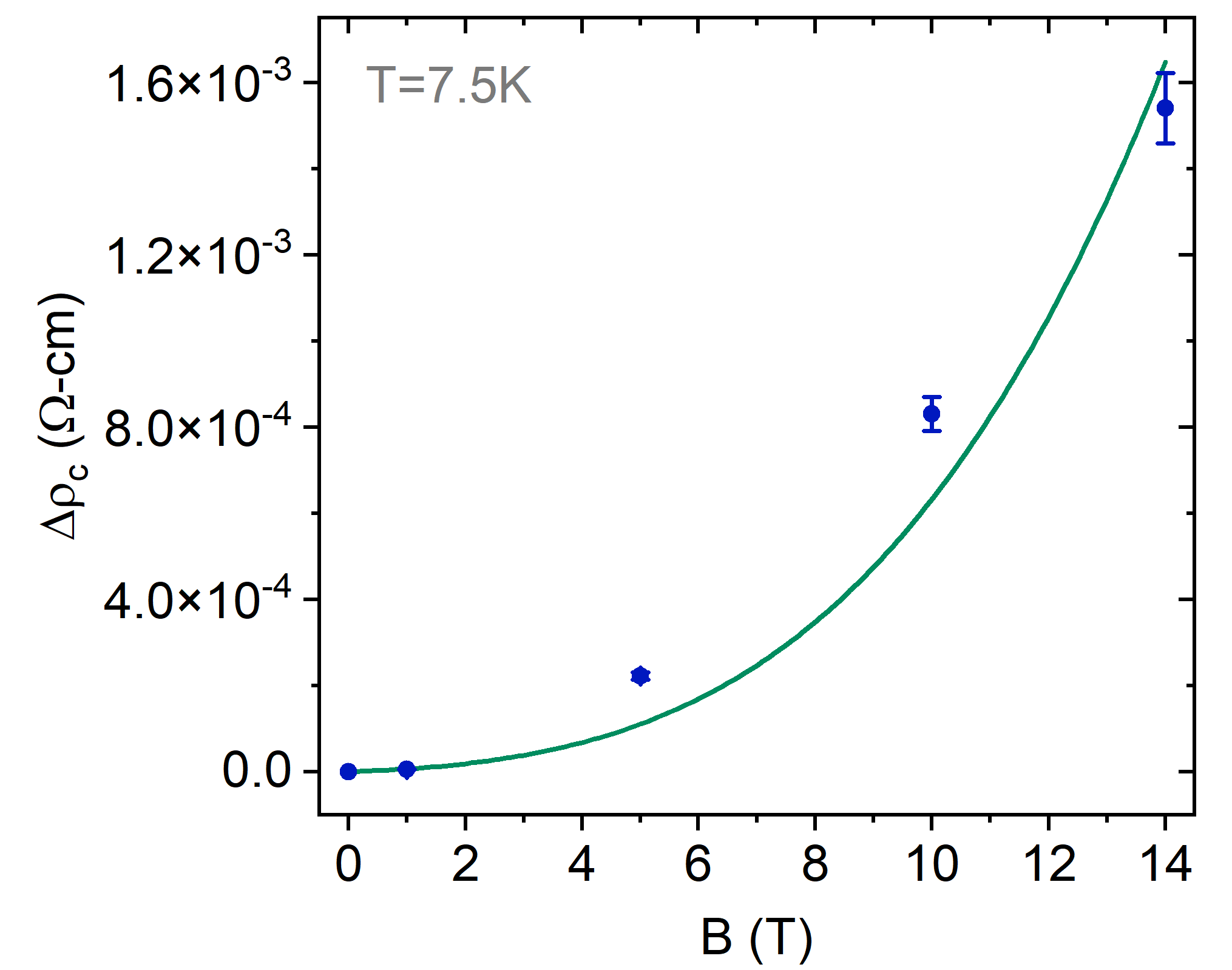}
	\caption{The dependence of amplitude of the $c$-axis resistivity oscillation on the magnitude of the in-plane magnetic field. The blue solid line is the theoretical prediction from Eq. (\ref{eq:rhoc}) and (\ref{eq:rhoc2}), and the orange solid dots are from the experimental data. Materials and methods for these original data are discussed in Appendix \ref{sec:MM}.}
	\label{fig:field2}
\end{figure}

Since the spin degree of freedom is irrelevant in the following discussion, we focus on a model of spinless Dirac fermions. In terms of the momentum label $k_z$ in the $c$ direction, the Hamiltonian can be written as
\begin{equation}
	\begin{split}
		& H_{\text{quasi}}=\sum_{\bm{k},k_z}\Psi^{\dagger}(\bm{k},k_z)\mathcal{H}(\bm{k},k_z)\Psi(\bm{k},k_z), \quad \Psi=\begin{pmatrix} \psi_+ \\ \psi_- \end{pmatrix}, \\
		& \mathcal{H}(\bm{k},k_z)=\begin{pmatrix}
		  	v_xk_x\sigma_x+v_yk_y\sigma_y & -t(1+e^{-\ri k_za_0})\mathbb{I} \\
		  	-t(1+e^{+\ri k_za_0})\mathbb{I} & v_yk_x\sigma_x+v_xk_y\sigma_y
		  \end{pmatrix},
	\end{split}
\end{equation}
where $\bm{k}=(k_x,k_y)$ is the in-plane momentum, $\pm$ labels the two layers in a unit cell and $a_0$ is the lattice constant in the $c$ direction. The anisotropy shown in Fig. \ref{fig:FS1} is characterized by $\gamma^2\equiv v_x/v_y \neq 1$. Now we apply an in-plane magnetic field in the $ab$ plane:
\begin{equation}
	\bm{B}=(B\cos\theta,B\sin\theta,0) ~~\Rightarrow~~\bm{A}=Bz(\sin\theta,-\cos\theta,0),
\end{equation}
where $\theta$ is the angle between the in-plane magnetic field and the stripe direction. It does not matter with respect to which layer in the unit cell $\theta$ is defined, since the system is invariant under a rotation by 90 degrees. The in-plane magnetic field appears in the Hamiltonian through
\begin{equation}
	(k_x,k_y) ~~\to~~ (k_x-q_ezB\sin\theta,k_y+q_ezB\cos\theta),
\end{equation}
where $q_e$ is the electron charge and $z=-\ri\rd_{k_z}$ acts a differential operator. The $c$-axis resistivity can be calculated via the Kubo formula:
\begin{equation}
	\begin{split}
		&\frac{1}{\rho_c^{n}}=-\frac{q_e^2t^2}{2\pi}\int\frac{\rd^2 k~\rd k_z\rd k'_z}{(2\pi)^4}\int \frac{\rd y}{2T}~\text{sech}^2\left( \frac{y}{2T} \right) \\
		& \times \operatorname{Tr}\Big{[} \left(e^{-\ri k_za_0+\ri k'_za_0}\right) \\
		& \quad \quad \quad \times \operatorname{Im}G^{R}_-(y;\bm{k};k_z,k'_z)\operatorname{Im}G^R_+(y;\bm{k};k'_z,k_z) \\
		& \quad \quad +\left(e^{\ri k_za_0-\ri k'_za_0}\right)\\
		& \quad \quad \quad \times \operatorname{Im}G^{R}_+(y;\bm{k};k_z,k'_z)\operatorname{Im}G^R_-(y;\bm{k};k'_z,k_z)\Big{]},
	\end{split}
\end{equation}
where the Green's functions  are defined within each layer in a unit cell. For a meaningful calculation, we introduce point disorders within each layer, which can be of two types for the Dirac fermions:
\begin{equation}
	\mathcal{H}_{\text{dis}}(\bm{r})=\begin{pmatrix}
		V_1(\bm{r})+V_2(\bm{r})\sigma_z & 0 \\ 0 & V_1(\bm{r})+V_2(\bm{r})\sigma_z
	\end{pmatrix},
\end{equation}
where the point disorders are short-range correlated:
\begin{equation}
	\overbar{V_i(\bm{r})}=0, ~ \overbar{V_i(\bm{r})V_j(\bm{r}')}=\Delta_i\delta_{ij}\delta^{(2)}(\bm{r}-\bm{r}'), ~ i,j=1,2.
\end{equation}
These point disorders can be caused by point defects and vortex cores within the Cu-O plane. The effect of the point disorders can be collectively characterized by a parameter $\beta\equiv (\Delta_1+\Delta_2)/(4v_xv_y)$, and we are assuming weak disorder $\beta\ll 1$. To the leading order in $t^2$, and under the condition of large anisotropy ($\gamma^2\ll 1$ or $\gamma^2\gg 1 $), the $c$-axis resistivity $\rho^n_c(\theta)$ can be approximated by the following expression:
\begin{equation}
\label{eq:rhoc}
	\begin{split}
		& \rho^n_c(\theta)=\frac{\rho_{c,0}}{g(\theta)+g(\pi/2-\theta)}, \\
		& g(\theta)=\sqrt{\frac{\frac{1}{\gamma^2}\cos^2\theta+\gamma^2\sin^2\theta}{\frac{1}{\gamma^2}\sin^2\theta+\gamma^2\cos^2\theta}} \\
		& \quad \quad \quad \times f\left(\xi\sqrt{\frac{1}{\gamma^2}\cos^2\theta+\gamma^2\sin^2\theta}\right),\\
		& f(x)=\int \rd y~\text{sech}^2y~\frac{y^2}{y^2+x^2},
	\end{split}
\end{equation}
where $\rho_{c,0}\propto \beta t^{-2}$ does not depend on $\theta$, and the parameters $\gamma,\xi$ are defined as
\begin{equation}
	\gamma^2=v_x/v_y, \quad \xi=q_eBa_0\sqrt{v_xv_y}/(2T).
\end{equation}
The result respects the symmetry under $\gamma\to 1/\gamma$, which is required since we have the freedom to name the directions. The fitting of Fig. \ref{fig:resis0} according to Eq. (\ref{eq:rhoc}) is shown in Fig. \ref{fig:resis1}.
It is clear that the $c$-axis resistivity varies periodically with the direction of the in-plane magnetic field. The period of the resistivity oscillation is $\pi/2$, and the minima of the resistivity lie at $k\pi/2$ while the maxima lie at $(2k+1)\pi/4$, where $k=0,1,2\cdots$. These features agree well with the experimental data. In plotting Fig. \ref{fig:resis1}, we assumed an average velocity $v=\sqrt{v_xv_y}$ equal to 0.1\% the speed of light, roughly the velocity of the Dirac fermions in graphene. Accordingly, to produce the observed amplitude of oscillation shown in Fig. \ref{fig:resis0}, we obtain a relatively large anisotropy $\gamma^2\approx 0.18$, or equivalently $\gamma^2\approx 5.7$. In contrast, as discussed in Appendix \ref{sec:quadratic}, the conventional fermions with quadratic yet anisotropic dispersion cannot produce the observed amplitude of oscillation shown in Fig. \ref{fig:resis0} with a realistic fermion mass. In this sense, the experimental data favor the possibility of Dirac fermions. By combining Eq. (\ref{eq:rhoc}) together with Eq. (\ref{eq:rhoc2}), we can obtain the dependence of the $c$-axis resistivity on the magnitude of the in-plane magnetic field for a fixed temperature. We show in Fig. \ref{fig:field} and \ref{fig:field2} the fitting of our result to the experimental data, which utilizes the anisotropy parameter $\gamma=0.42$ obtained previously. The consistency is reasonably good given our simple model of anisotropy and approximate functional form for large anisotropy.

\begin{figure}[htp!]
	\centering
	\includegraphics[width=0.95\linewidth]{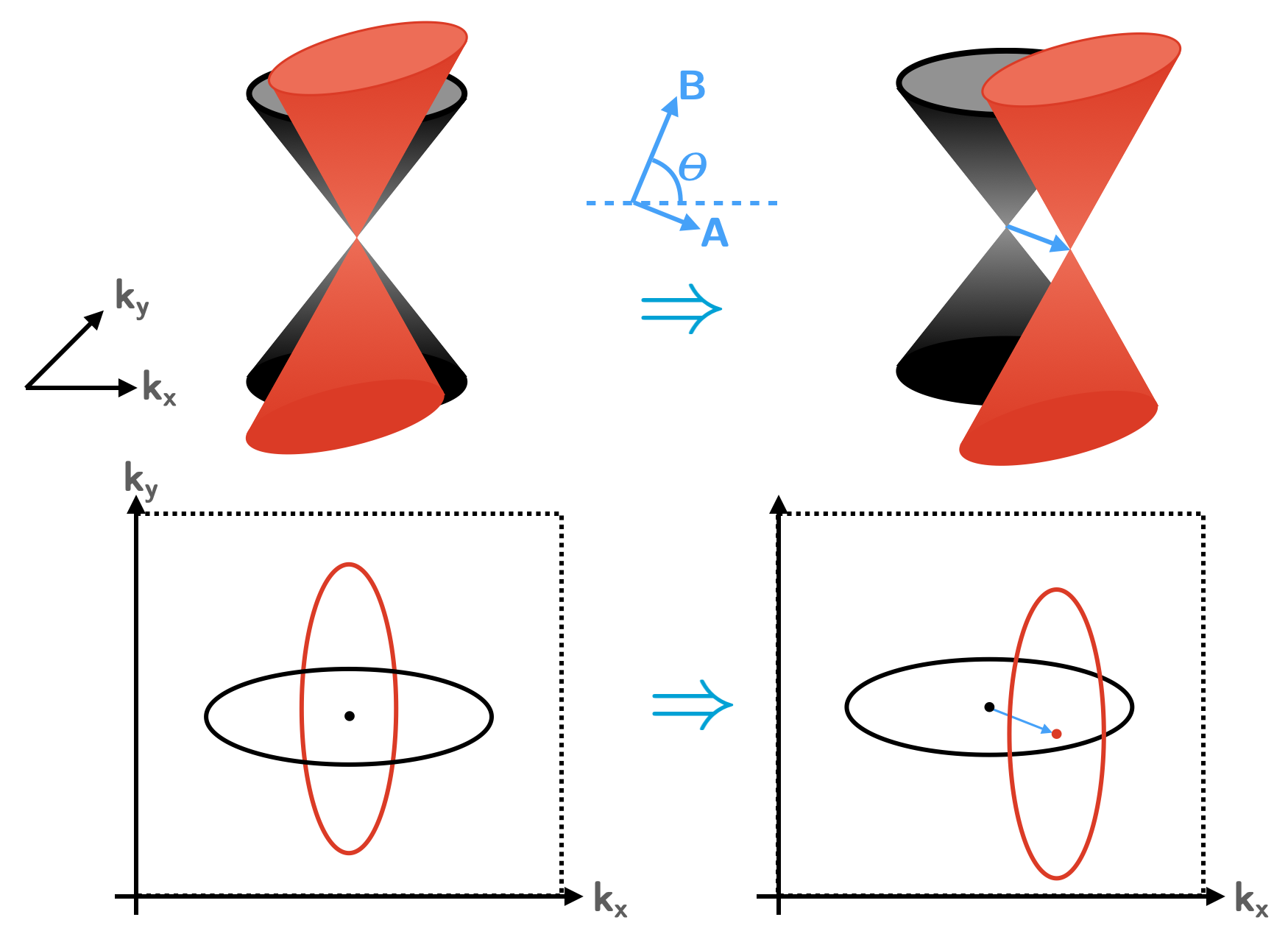}
	\caption{The schematic illustration of the mechanism of the oscillation observed in the $c$-axis resistivity in presence of an in-plane magnetic field. The tips of the Dirac cones in nearest neighboring layers are shifted by the in-plane magnetic field. The magnitude of the shift is fixed by the strength of the magnetic field and the lattice constant, while the direction of the shift varies with the direction of the magnetic field. If there is anisotropy, such a shift will produce a varying overlap between the cross sections, and thus a varying $c$-axis resistivity.}
	\label{fig:shift}
\end{figure}

The angular dependence of the $c$-axis resistivity in the simplified model described above can be understood as follows. In order to conduct along $c$ axis, the Dirac fermions have to hop between layers. From one layer to another, the probability of such hopping is proportional to the amount of overlapping between the two cross sections of the Dirac cones determined by the temperature, as shown in Fig. \ref{fig:FS1}. The in-plane magnetic field has the effect to shift the cross sections in different layers by different amount along the direction perpendicular to the in-plane magnetic field, thus changing their overlap, as shown in Fig. \ref{fig:shift}. If the Dirac cone in each layer is isotropic, such shifting will also be isotropic, then the overlapping between the cross sections and consequently the hopping is also isotropic, resulting in an angle-independent $c$-axis resistivity. Otherwise, if there is anisotropy in the Dirac cone, the amount of overlapping between the cross sections changes with the shifting direction, resulting in an angle-dependent hopping and $c$-axis resistivity. The above argument is illustrated schematically in Fig. \ref{fig:shift}. The periodicity in the angular dependence is fixed by the rotational symmetry - in each layer, the stripe configuration gives us a $C_2$ symmetry, while in each unit cell, the orthogonal stripe directions of the neighboring layers enhance the symmetry to $C_4$. As a result, we will have a four-fold periodicity, namely the $\pi/2$ period in the $c$-axis resistivity. As a bonus point, the anisotropy in the Dirac cone as well as the hopping along $c$ axis relaxes the quasi-2D nature of the system inherited from the layer structure, to the extent that the weak localization effect is negligible, leaving us a finite $c$-axis resistivity.

\section{Discussion}
\label{sec:disscussion}

In the current paper, we have argued that  the recent angle-resolved $c$-axis resistivity measurements of the stripe-ordered LBCO, together with temperature dependence,  imply the existence of two weakly coupled subsystems containing superconducting pairs of different symmetry. Pairs of the first type are located on charge stripes and are confined to $ab$ planes with the peculiar PDW configuration first proposed in \cite{theory}, such that the interlayer Josephson coupling is nullified. The pairs of the other type must have a different symmetry, they are most likely conventional $d$-wave pairs, whose presence is due to doping inhomogeneities. The one-dimensional character of the $c$-axis transport indicates that these pairs must traverse the sample along narrow channels along the $c$ direction without interruption, likely formed by columnar defects. The layer decoupling can be destroyed by defects in the charge-stripe order. Then, with decreasing temperature, the two subsystems merge into one and first go through a glassy transition to a dissipationless state, and then evolve into a Meissner state.

The phenomenological model of the two-subsystem picture consists of preformed pairs moving in the charge stripes, and pairing of quasiparticles confined in the columnar defects. It can provide a good, qualitative explanation of the temperature dependence of the experimental data. In the mean time, the observed dependence of the $c$-axis resistivity on the direction of the in-plane magnetic field can be explained by introducing anisotropy of the normal state quasiparticles compatible with the peculiar in-plane stripe order, and the experimental data favor the possibility of the quasiparticles being Dirac fermions (which we will discuss further at the end of this section). As a consequence, when we apply a magnetic field strong enough to destroy the coherence between the preformed pairs, and free the Dirac fermions to propagate in three dimensions, we will obtain a resistive state in presence of weak disorder. Such a resistive state has a zero Hall response due to the particle-hole symmetry of the Dirac fermions, which is consistent with the experimental findings \cite{tranquada}. This gives support to the quasiparticle picture over the Bose metal picture about the nature of such a resistive state.

The Dirac fermion picture favored by the experimental data as proposed in this paper is unusual, so we emphasize the logic behind our reasoning here. To explain the magneto-resistivity, we have two options at hand, one is a quasiparticle mechanism, and the other is a Bose metal. We can rule out the later option since the incoherent pairs would not react to the direction of the magnetic field. As a consequence there would be no oscillation of the magneto-resistivity as seen in Fig. \ref{fig:resis0}. As far as the quasiparticle mechanism is concerned, the first candidate is the conventional quasiparticles with quadratic dispersion. However, in Appendix \ref{sec:quadratic} we have shown that in order to explain the experimental data, the quasiparticle mass must be vanishingly small. This brings us to the next candidate on the list, which is the Dirac fermions. We have shown in the main text that the Dirac fermions produce a surprising agreement with the experimental data (the oscillation pattern with respect to the direction of the in-plane magnetic field, and the oscillation amplitude with respect to the strength of the in-plane magnetic field), given the simplicity of our phenomenology. Therefore, our result provides a minimal phenomenological model that is consistent with the experimental data. Although we are forced to introduce the Dirac quasiparticles by the weight of the experimental data, we admit that at present we cannot suggest any microscopic justification for this hypothesis. To date, Dirac fermions are found in topological insulators and semimetals like graphene. For $d$-wave superconducting cuprates, spinons with linear dispersion are proposed for the undoped system \cite{PhysRevB.37.3774}, and Bogoliubov quasiparticles (effectively of spin $1/2$ and charge $0$) can present a Dirac fermion type dispersion along the nodal direction of the Brillouin zone \cite{doi:10.1146/annurev-conmatphys-031113-133841}, but the normal state quasiparticles (of spin $1/2$ and charge $e$) with Dirac fermion type dispersion still evade direct or indirect experimental observations. To our knowledge, the only exception is the observation of a closely related type-II Dirac fermion in cuprates with a different chemical substitution, La$_{1.77}$Sr$_{0.23}$CuO$_4$, where the same Cu and O bands are involved and the observed tilted linear dispersion is attributed to the tight-binding band structure \cite{LSCO}. Therefore, we hope that our work may encourage theoretical efforts in the direction of Dirac fermions in the stripe-ordered LBCO.

The phenomenological theory discussed in this paper has relations with other theories previously proposed. The theoretical description in \cite{tsvelik19} provides a possible microscopic mechanism for the preformed pairs moving on the charge stripes and quasiparticles hopping between layers, using a strongly correlated Kondo lattice model, but the quasiparticles are conventional fermions with quadratic dispersion. On the other hand, Dirac fermions have been proposed \cite{PhysRevB.93.205116,PhysRevB.95.045118} to explain the magnetically induced resistive state in disordered superconducting films, but the Dirac fermions are of the composite type. Thus, a consistent microscopic mechanism for the phenomenological theory discussed in this paper requires further investigations. On the experimental side, a definitive experimental evidence should come from the angle-resolved photoemission spectroscopy (ARPES) identified with the normal state band structure, as has been done for La$_{1.77}$Sr$_{0.23}$CuO$_4$. The position of the Dirac point should be tunable by changing the chemical substitution and the doping level, which is also subject to experimental observations. In addition to the Dirac fermions, the formation mechanism, the geometric property, and the distribution of the narrow superconducting channels that host the Dirac fermions also call for a further investigation. Other experimental observations such as thermal transport measurements are need to obtain a full understanding of these detailed properties.

\section{Acknowledgement}
We are grateful to John Tranquada for his suggestion of the possible anisotropy mechanism and Dragana Popovic for extensive discussions and valuable suggestions. This work was supported by Office of Basic Energy Sciences, Material Sciences and Engineering Division, U.S. Department of Energy (DOE) under Contracts No. DE-SC0012704. This paper was written in part while AMT was a visitor at KITP, which is supported by by the National Science Foundation under Grant No. NSF PHY-1748958.

\appendix
\numberwithin{equation}{section}

\section{Quasiparticles with Linear Dispersion}
\label{sec:linear}

Here we show details of the calculation of the $c$-axis normal conductivity $\sigma_c$ due to hopping of quasiparticles with linear dispersion (namely the Dirac fermions), using the simplified model introduced in the main text. The inverse of $\sigma_c$ will be the $c$-axis normal resistivity $\rho^n_c$ discussed in the main text.

\subsection{Green's Function for Linear Dispersion}

The Hamiltonian for the Dirac fermions is
\begin{equation}
	\begin{split}
		& H_{\text{quasi}}=\sum_{\bm{k},k_z}\Psi^{\dagger}(\bm{k},k_z)\mathcal{H}(\bm{k},k_z)\Psi(\bm{k},k_z), ~ \Psi=\begin{pmatrix} \psi_+ \\ \psi_- \end{pmatrix}, \\
		& \mathcal{H}(\bm{k},k_z)=\begin{pmatrix}
		  	v_xk_x\sigma_x+v_yk_y\sigma_y & -t(1+e^{-\ri k_za_0})\mathbb{I} \\
		  	-t(1+e^{+\ri k_za_0})\mathbb{I} & v_yk_x\sigma_x+v_xk_y\sigma_y
		  \end{pmatrix},
	\end{split}
\end{equation}
where $\bm{k}=(k_x,k_y)$, $\pm$ labels the two layers in a unit cell, and $a_0$ is the lattice constant in the $c$ direction. In presence of the in-plane magnetic field, we need to perform the following substitution:
\begin{equation}
\label{eq:subs}
	\begin{split}
		& k_x \to \tilde{k}_x=k_x-q_ezB\sin\theta\\
		& k_y \to \tilde{k}_y= k_y+q_ezB\cos\theta,
	\end{split} 
\end{equation}
where $z=-\ri \rd_{k_z}$ in the momentum representation. The Matsubara Green's function $\hat{G}\equiv -\lrangle{\Psi\Psi^{\dagger}}$ is determined by the following equation:
\begin{equation}
	\begin{split}
		\left[\ri\omega_n-\mathcal{H}(\bm{k};k_z)\right]\hat{G}(\ri\omega_n;\bm{k};k_z,k'_z)=\delta(k_z-k'_z)\begin{pmatrix}\mathbb{I} & 0 \\ 0 & \mathbb{I}	 \end{pmatrix}.
	\end{split}
\end{equation}
The matrix Green's function $\hat{G}$ can be partitioned as
\begin{equation}
	\begin{split}
		& \hat{G}=\begin{pmatrix} G_{+} & F \\ \bar{F} & G_{-} \end{pmatrix}, \\
		& \begin{cases}
			G_+=-\lrangle{\psi_+\psi^{\dagger}_+}, & G_-=-\lrangle{\psi_-\psi^{\dagger}_-} \\
			F=-\lrangle{\psi_+\psi^{\dagger}_-}, & \bar{F}=-\lrangle{\psi_-\psi^{\dagger}_+}
		\end{cases}.
	\end{split}
\end{equation}
To evaluate the conductivity to leading order in $t$, we can just keep the zeroth order Green's functions:
\begin{equation}
\label{eq:linearzero}
	\begin{split}
		& F^{(0)}(\ri \omega_n;\bm{k};k_z,k'_z)=0, \quad \bar{F}^{(0)}(\ri \omega_n;\bm{k};k_z,k'_z)=0,\\
		& G^{(0)}_+(\ri\omega_n;\bm{k};k_z,k'_z)\\
		=&-\frac{1}{2\pi}\int\rd z\frac{\ri\omega_n+v_x\tilde{k}_x\sigma_x+v_y\tilde{k}_y\sigma_y}{\omega^2_n+m^2+v^2_x\tilde{k}^2_x+v^2_y\tilde{k}^2_y}e^{\ri(k_z-k'_z)z},\\
		& G^{(0)}_-(\ri\omega_n;\bm{k};k_z,k'_z)\\
		=&-\frac{1}{2\pi}\int\rd z\frac{\ri\omega_n+v_y\tilde{k}_x\sigma_x+v_x\tilde{k}_y\sigma_y}{\omega^2_n+m^2+v^2_y\tilde{k}^2_x+v^2_x\tilde{k}^2_y}e^{\ri(k_z-k'_z)z},
	\end{split}
\end{equation}
and the physical Green's function $G^R_{\pm}$ can be obtained by the analytic continuation $\ri \omega_n\to \omega+\ri 0^{+}$.
Now we introduce the point disorders within each layer mentioned in the main text:
\begin{equation}
	\begin{split}
		& \mathcal{H}_{\mathrm{dis}}(\boldsymbol{r})=\left(\begin{array}{cc}
			V_{1}(\boldsymbol{r})+V_{2}(\boldsymbol{r}) \sigma_{z} & 0 \\
			0 & V_{1}(\boldsymbol{r})+V_{2}(\boldsymbol{r}) \sigma_{z}
		\end{array}\right),\\
		& \overline{V_{i}(\boldsymbol{r})}=0, \quad \overline{V_{i}(\boldsymbol{r}) V_{j}\left(\boldsymbol{r}^{\prime}\right)}=\Delta_{i} \delta_{i j} \delta^{(2)}\left(\boldsymbol{r}-\boldsymbol{r}^{\prime}\right),
	\end{split}
\end{equation}
where $i,j$ takes the values $1,2$.
The point disorders introduce a self-energy to the Dirac fermions as
\begin{equation}
	\begin{split}
		\Sigma^R(\omega;\bm{k};k_z,k'_z)=&\Delta_1 \int\frac{\rd^2k}{(2\pi)^2}G^R_{\pm}(\omega;\bm{k};k_z,k'_z)\\
		+&\Delta_2 \int\frac{\rd^2k}{(2\pi)^2}\sigma_zG^R_{\pm}(\omega;\bm{k};k_z,k'_z)\sigma_z.
	\end{split}
\end{equation}
The calculation of $\Sigma^R$ involves the following integral, which can be approximated as \cite{doi:10.1143/JPSJ.67.2421}
\begin{equation}
\label{eq:strategy}
	\begin{split}
		&\int\frac{\rd^2k}{(2\pi)^2}\frac{\omega+\ri 0^++v_x\tilde{k}_x\sigma_x+v_y\tilde{k}_y\sigma_y}{(\omega+\ri 0^+)^2-\left(v^2_x\tilde{k}^2_x+v^2_y\tilde{k}^2_y\right)}\\
		\approx &\frac{A}{2(2\pi)^2}\int_{-\infty}^{\infty}\rd x\lrangle{\frac{\omega+\ri 0^++v_x\tilde{k}_x\sigma_x+v_y\tilde{k}_y\sigma_y}{(\omega+\ri 0^+)^2-x}}_{\omega},\\
		=&\frac{A}{2(2\pi)^2}\int_{-\infty}^{\infty}\rd x \frac{\omega+\ri 0^+}{(\omega+\ri 0^+)^2-x}=-i\frac{|\omega|}{4v_xv_y},
	\end{split}
\end{equation}
where the average is over the cross section $v^2_x\tilde{k}^2_x+v^2_y\tilde{k}^2_y=\omega^2$, and $A=\nu(|\omega|)/|\omega|=2\pi/(v_xv_y)$ with $\nu(|\omega|)$ being the density of states. In the end, the disorder averaged Green's function can be represented as
\begin{equation}
\label{eq:disGR}
	\begin{split}
		& \overbar{G^R_+(\omega;\bm{k};k_z,k'_z)}\\
		=&-\frac{1}{2\pi}\int\rd z\frac{\omega+\ri \Gamma+v_x\tilde{k}_x\sigma_x+v_y\tilde{k}_y\sigma_y}{-(\omega+\ri \Gamma)^2+v^2_x\tilde{k}^2_x+v^2_y\tilde{k}^2_y}e^{\ri(k_z-k'_z)z}, \\
		& \overbar{G^R_-(\omega;\bm{k};k_z,k'_z)}\\
		=&-\frac{1}{2\pi}\int\rd z\frac{\omega+\ri \Gamma+v_y\tilde{k}_x\sigma_x+v_x\tilde{k}_y\sigma_y}{-(\omega+\ri \Gamma)^2+v^2_y\tilde{k}^2_x+v^2_x\tilde{k}^2_y}e^{\ri(k_z-k'_z)z},
	\end{split}
\end{equation}
where $\Gamma(\omega)=|\omega|(\Delta_1+\Delta_2)/(4v_xv_y)$. In the following calculations, we assume that $\Gamma(\omega) \ll |\omega|$, which means a weak disorder $\beta\equiv (\Delta_1+\Delta_2)/(4v_xv_y)\ll 1$.

\subsection{Kubo Formula for $c$-axis Conductivity} 

We focus on the DC conductivity obtained as setting $\bm{k}\to \bm{0}$ and taking the limit $\omega\to 0$ in the end. The current operator with $\bm{k}\to \bm{0}$ is
\begin{equation}
	\begin{split}
		&J_z(\bm{0},\omega)=q_et\sum_{\bm{k}}\left[\Psi^{\dagger}(\omega;\bm{k},k_z)\Sigma_z(k_z)\Psi(\omega;\bm{k},k_z)\right],\\
		& \Sigma_z(k_z)=\begin{pmatrix}
			0 & \ri e^{-\ri k_z a_0}\mathbb{I} \\ -\ri e^{\ri k_za_0}\mathbb{I} & 0
		\end{pmatrix}.
	\end{split}
\end{equation}
Then the $c$-axis conductivity can be derived from the Kubo formula
\begin{equation}
	\operatorname{Re}\sigma_c(\omega)=\frac{1-e^{-\beta\omega}}{2\omega V}\int_{-\infty}^{\infty}\rd t~e^{\ri \omega t}\lrangle{J^{\dagger}_z(\bm{0},t)J_z(\bm{0},0)},
\end{equation}
where $V=V_{2D}V_z$ is the size of the system. In the leading order ($\sigma_c\propto t^2$), we obtain
\begin{equation}
	\sigma_c(\omega)=\frac{\ri}{\omega}q_e^2t^2\Pi_{zz}(\omega),
\end{equation}
where $\Pi_{zz}(\omega)$ is obtained from its Matsubara representation $\Pi_{zz}(\ri \Omega_n)$:
\begin{equation}
	\begin{split}
		&\Pi_{zz}(\ri \Omega_n)\\
		=&\frac{1}{\pi^2 V}\sum_{\bm{k}}\sum_{k_z,k'_z}\iint \rd\xi\rd\xi'\frac{\tanh(\xi/2T)-\tanh(\xi'/2T)}{i\Omega_n+\xi-\xi'}\\
		& \times \operatorname{Tr}\Big{[} \left(e^{-\ri k_z a_0+\ri k'_za_0}\right) \\
		& \quad \quad \quad \times \operatorname{Im}G^{R}_-(\xi;\bm{k};k_z,k'_z)\operatorname{Im}G^R_+(\xi';\bm{k};k'_z,k_z) \\
		& \quad \quad +\left(e^{\ri k_z a_0-\ri k'_za_0}\right)\\
		& \quad \quad \quad \times \operatorname{Im}G^{R}_+(\xi;\bm{k};k_z,k'_z)\operatorname{Im}G^R_-(\xi';\bm{k};k'_z,k_z)\Big{]}.
	\end{split}
\end{equation}
Then the $c$-axis conductivity in the DC limit $\omega\to 0$ is
\begin{equation}
\label{eq:caxis0}
	\begin{split}
		\sigma_c=&-\frac{q_e^2t^2}{\pi V}\sum_{\bm{k}}\sum_{k_z,k'_z}\int\frac{\rd y}{2T}~\text{sech}^2\left(\frac{y}{2T} \right)\\
		& \times \operatorname{Tr} \Big{[} \left(e^{-\ri k_z a_0+\ri k'_z a_0}\right) \\
		& \quad \quad \quad \times \operatorname{Im}G^{R}_-(y;\bm{k};k_z,k'_z)\operatorname{Im}G^R_+(y;\bm{k};k'_z,k_z) \\
		& \quad \quad +\left(e^{\ri k_z a_0-\ri k'_z a_0}\right) \\
		& \quad \quad \quad \times \operatorname{Im}G^{R}_+(y;\bm{k};k_z,k'_z)\operatorname{Im}G^R_-(y;\bm{k};k'_z,k_z)\Big{]},
	\end{split}
\end{equation}
and we need to perform the disorder average for the expression in Eq. (\ref{eq:caxis0}). Under the weak disorder condition $\beta\ll 1$, we can ignore the vertex correction and replace the individual Green's function $G^{R}_{\pm}$ with its disorder average $\overbar{G^R_{\pm}}$ determined in Eq. (\ref{eq:disGR}).

\subsection{Calculation of the $c$-axis Conductivity} 
We consider the evaluation of Eq. (\ref{eq:caxis0}) with $G^{R}_{\pm}$ replaced with $\overbar{G^R_{\pm}}$. The integration over $k_z,k_z'$ can be carried out straightforwardly, while the integration over $\bm{k}$ can be carried out using the strategy shown in Eq. (\ref{eq:strategy}). Finally, we obtain
\begin{equation}
\label{eq:sc}
	\sigma_{c}=\frac{q_e^2t^2}{2\pi v_xv_y }\int \frac{\rd y}{2T}~\text{sech}^2\left( \frac{y}{2T} \right)\left[ I(y,\theta)+I^*(y,\theta) \right]
\end{equation}
where the function $I(y)$ has the following expression:
\begin{equation}
	\begin{split}
		I(y>0,\theta)=&\left(\lrangle{\frac{y^2+\Gamma^2+v_xv_y\left(k_xk'_x+k_yk'_y\right)}{4y\Gamma-iF_+}}_+\right.\\
		+&\left.\lrangle{\frac{y^2+\Gamma^2+v_xv_y\left(k_xk'_x+k_yk'_y\right)}{4y\Gamma+iF_-}}_-\right), \\
		I(y<0,\theta)=&\left(\lrangle{\frac{y^2+\Gamma^2+v_xv_y\left(k_xk'_x+k_yk'_y\right)}{4|y|\Gamma+iF_+}}_+\right.\\
		+&\left.\lrangle{\frac{y^2+\Gamma^2+v_xv_y\left(k_xk'_x+k_yk'_y\right)}{4|y|\Gamma-iF_-}}_-\right).
	\end{split}
\end{equation}
The subscript $\pm$ denotes the two layers within a unit cell, and the average is parameterized by $\alpha\in[0,2\pi)$ according to the following rules:
\begin{widetext}
\begin{equation}
	\begin{split}
		+: &  \quad k_x=\frac{|y|\cos\alpha}{v_x}, \quad k_y=\frac{|y|\sin\alpha}{v_y}; \quad k'_x=-q_eBa_0\sin\theta+\frac{|y|\cos\alpha}{v_x}, \quad k'_y=q_eBa_0\cos\theta+\frac{|y|\sin\alpha}{v_y}, \\
		& \quad F_+=v^2_y\left(-q_eBa_0\sin\theta+|y|\cos\alpha/v_x \right)^2+v^2_x\left(q_eBa_0\cos\theta+|y|\sin\alpha/v_y\right)^2-y^2, \\
		-: &  \quad k_x=q_eBa_0\sin\theta+\frac{|y|\cos\alpha}{v_y}, \quad k_y=-q_eBa_0\cos\theta+\frac{|y|\sin\alpha}{v_x}; \quad k'_x=\frac{|y|\cos\alpha}{v_y}, \quad k'_y=\frac{|y|\sin\alpha}{v_x}, \\
		& \quad F_-=v^2_x\left(q_eBa_0\sin\theta+|y|\cos\alpha/v_y \right)^2+v^2_y\left(-q_eBa_0\cos\theta+|y|\sin\alpha/v_x\right)^2-y^2.
	\end{split}
\end{equation}
\end{widetext}
Introducing the following parameters:
\begin{equation}
	\begin{split}
		& \epsilon_0\equiv vq_eBa_0, \quad v_x=v\gamma, \quad v_y=v/\gamma,\\
		& \beta=(\Delta_1+\Delta_2)/(4v^2),
	\end{split}
\end{equation}
we can express $I(y,\theta)+I^*(y,\theta)=J(y,\theta)+J(y,\pi/2-\theta)$ as
\begin{equation}
	\begin{split}
		& J(y,\theta) =\lrangle{\frac{8\beta y^2\left[(1+\beta^2)y^2+P \right]}{(4\beta y^2)^2+Q^2}}_{\alpha \in [0, 2\pi)},\\
		& P(\alpha)=\left(\frac{y^2\cos^2\alpha}{\gamma^2}+\gamma^2y^2\sin^2\alpha+\gamma\epsilon_0|y|\cos\theta\sin\alpha\right. \\
		& \quad \quad \quad \quad \left.-\frac{\epsilon_0|y|\sin\theta\cos\alpha}{\gamma}\right), \\
		& Q(\alpha)=\frac{1}{\gamma^2}\left(\frac{|y|\cos\alpha}{\gamma}-\epsilon_0\sin\theta \right)^2\\
		& \quad \quad \quad +\gamma^2\left(\gamma |y|\sin\alpha+\epsilon_0\cos\theta \right)^2-y^2. \\
	\end{split}
\end{equation}
An important feature of $\sigma_c$ expressed in the above formulas is that it is symmetric under $\gamma\to 1/\gamma$, since the system is invariant under rotation by 90 degrees, or in another word, we have the freedom to define $x,y$-axis in neighboring layers. The analytical and numerical evaluation of the average over $\alpha$ presents obstacles due to the singular behavior of the integrand. As a practical approximation, we work in the limit $\gamma \ll 1$ and $\beta\ll 1$, and take into account only the dominating contributions near the singular points of the integrand. The singular points are determined by
\begin{equation}
	Q\approx 0 \quad \Rightarrow \quad \cos\alpha_{1,2} \approx \frac{\gamma \epsilon_0\sin\theta}{|y|}, \quad \alpha_1+\alpha_2=\pi,
\end{equation}
and the function $J(y,\theta)$ is approximated as
\begin{equation}
	\begin{split}
		J(y,\theta)\approx & \frac{1}{2\pi}\int \rd \alpha\frac{8\beta y^2\left[y^2+P(\alpha_1)\right]}{(4\beta y^2)^2+[Q'(\alpha_1)]^2(\alpha-\alpha_1)^2}\\
		+&\frac{1}{2\pi}\int \rd \alpha\frac{8\beta y^2\left[y^2+P(\alpha_2)\right]}{(4\beta y^2)^2+[Q'(\alpha_2)]^2(\alpha-\alpha_2)^2} \\
		\approx & \frac{1}{\gamma^6}\left|\frac{\cos\theta}{\sin\theta} \right|\frac{y^2}{y^2+\epsilon^2_0\cos^2\theta/\gamma^2}.
	\end{split}
\end{equation}
Substituting this into Eq. (\ref{eq:sc}), we obtain
\begin{equation}
\label{eq:result0}
\begin{split}
	& \sigma_c(\theta)\approx\frac{q^2_et^2}{2\pi\beta v_xv_y}\frac{1}{\gamma^6}\left[K(\theta)+K(\pi/2-\theta)\right]\\
	& K(\theta)=\left|\frac{\cos\theta}{\sin\theta} \right|\int  \frac{\rd y}{2T}~\text{sech}^2\left( \frac{y}{2T} \right)~\frac{y^2}{y^2+\epsilon_0^2\cos^2\theta/\gamma^2}.
\end{split}
\end{equation}
This approximate result misses the information from the opposite limit $\gamma\gg 1$, which can be cured by performing a symmetrization according to the required symmetry under $\gamma\to 1/\gamma$. Finally, we obtain
\begin{equation}
\label{eq:result1}
	\begin{split}
		& \sigma_c(\theta)\approx \frac{q^2_et^2}{2\pi\beta v_xv_y}\left(\gamma^2+\frac{1}{\gamma^2}\right)^3\left[g(\theta)+g(\pi/2-\theta) \right], \\
		& g(\theta)=\sqrt{\frac{(1/\gamma^2)\cos^2\theta+\gamma^2\sin^2\theta}{(1/\gamma^2)\sin^2\theta+\gamma^2\cos^2\theta}}\\
		& \quad \quad \quad \times f\left(\xi\sqrt{(1/\gamma^2)\cos^2\theta+\gamma^2\sin\theta}\right),\\
		& f(x)=\int \rd y~\text{sech}^2y~\frac{y^2}{y^2+x^2},
	\end{split}
\end{equation}
where we have introduced the parameter $\xi=\epsilon_0/(2T)$. The fitting of Eq. (\ref{eq:result1}) to the experimental data is shown in the main text, and we obtain $\gamma^2=v_x/v_y\approx 0.18$ for the averaged velocity equal to 0.1\% the speed of light.

\section{Quasiparticles with Quadratic Dispersion}
\label{sec:quadratic}

Here we discuss the $c$-axis normal conductivity $\sigma_c$ due to hopping of quasiparticles with quadratic dispersion, where the inverse of $\sigma_c$ will be the $c$-axis normal resistivity $\rho^n_c$. We will show the reason why such conventional quasiparticles with quadratic dispersion fail to explain the experimental data.

\subsection{Green's Function for Quadratic Dispersion} The Hamiltonian for the conventional fermions is
\begin{equation}
	\begin{split}
		& H_{\text{quasi}}=\sum_{\bm{k},k_z}\Psi^{\dagger}(\bm{k},k_z) \mathcal{H}(\bm{k},k_z)\Psi(\bm{k}	,k_z), \quad  \Psi=\begin{pmatrix}
			\psi_+ \\ \psi_- \end{pmatrix},\\
		& \mathcal{H}=\begin{pmatrix}
				\epsilon_+(\bm{k}) & -t(1+e^{-\ri k_za_0})\\
				-t(1+e^{\ri k_za_0}) & \epsilon_-(\bm{k})
			\end{pmatrix},\\
		& \epsilon_+(k_x,k_y)=\epsilon_-(k_y,k_x)=\frac{k^2_x}{2m_x}+\frac{k^2_y}{2m_y}-\mu,
	\end{split}
\end{equation}
where in presence of the in-plane magnetic field, we need to perform the substitution shown in Eq. (\ref{eq:subs}). The Matsubara Green's function $\hat{G}\equiv -\lrangle{\Psi\Psi^{\dagger}}$ is determined by
\begin{equation}
	\left[\ri\omega_n-\mathcal{H}(\bm{k},k_z)\right]\hat{G}(i\omega_n;\bm{k};k_z,k'_z)=\delta(k_z-k'_z)\mathbb{I},
\end{equation}
where the matrix Green's function again can be partitioned as
\begin{equation}
	\hat{G}=\begin{pmatrix}
			G_+ & F \\
			\bar{F} & G_-
		\end{pmatrix},
\end{equation}
and to evaluate the conductivity to leading order in $t$, we can just keep the zeroth order Green's functions:
\begin{equation}
\label{eq:zero}
	\begin{split}
		& F^{(0)}(\ri \omega_n;\bm{k};k_z,k'_z)=0, \quad \bar{F}^{(0)}(\ri \omega_n;\bm{k};k_z,k'_z)=0,\\
		& G^{(0)}_+(\ri \omega_n;\bm{k};k_z,k'_z)=\frac{1}{2\pi}\int \rd z\frac{1}{\ri \omega_n-\epsilon_+(\bm{k},z)}e^{\ri(k_z-k'_z)z}, \\
		& G^{(0)}_-(\ri \omega_n;\bm{k};k_z,k'_z)=\frac{1}{2\pi}\int \rd z\frac{1}{\ri \omega_n-\epsilon_-(\bm{k},z)}e^{\ri(k_z-k'_z)z}.
	\end{split}
\end{equation}
where $\epsilon_{\pm}(\bm{k},z)\equiv\epsilon_{\pm}(k_x-q_ezB\sin\theta,k_y+q_ezB\cos\theta)$. We then take into account the effect of the disorder within each layer:
\begin{equation}
	\overbar{V}=0, \quad \overbar{V(\bm{r})V(\bm{r}')}=\Delta\delta^{(2)}(\bm{r}-\bm{r}').
\end{equation}
Consequently, the disorder-averaged physical Green's function can be presented as
\begin{equation}
\label{eq:RA}
	\overbar{G^R_{\pm}(\omega;\bm{k};k_z,k'_z)}=\frac{1}{2\pi}\int\rd z\frac{e^{\ri (k_z-k'_z)z}}{\omega+\ri /(2\tau)-\epsilon_{\pm}(\bm{k},z)},
\end{equation}
and $\tau$ is the elastic scattering relaxation time, which can be taken as a constant $\tau=1/(2\pi\nu \Delta)$ with $\nu$ being the constant density of states in 2D, if we consider low enough temperatures.

\subsection{Calculation of the $c$-axis Conductivity} 

\begin{figure}[htp!]
	\includegraphics[width=0.9\linewidth]{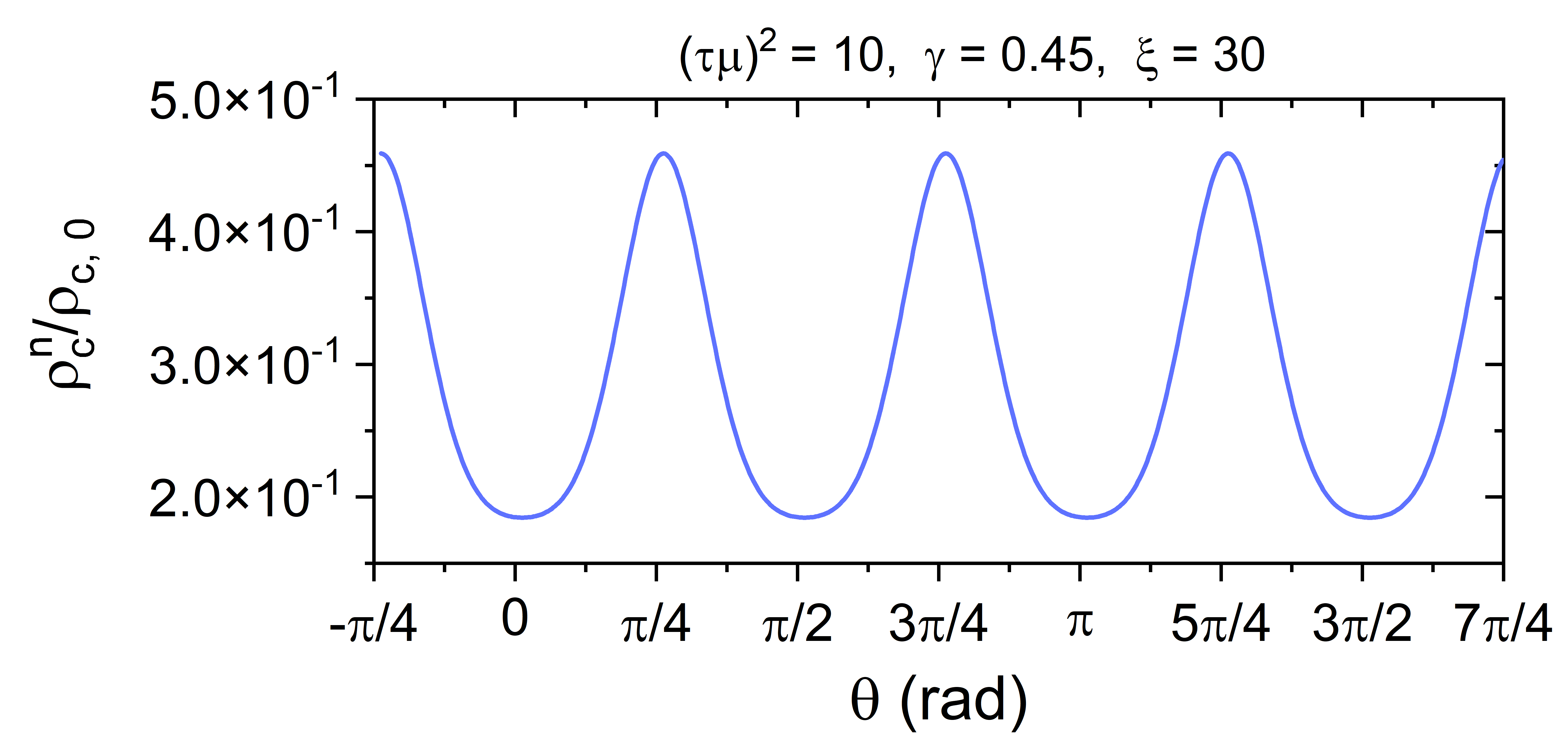}
	\caption{The schematic plot of the $c$-axis resistivity according to Eq. (\ref{eq:Drude}) and Eq. (\ref{eq:Drude1}), in correspondence to Fig. 4 in the main text.}
	\label{fig:resisqu}
\end{figure}

We can still use the Kubo formula shown in Eq. (\ref{eq:caxis0}), where the trace operation becomes trivial. In case of weak disorder such that the vertex correction can be ignored, we just replace the Green's function $G^R_{\pm}$ with the disorder-averaged ones $\overbar{G^R_{\pm}}$. Similar in spirit to the strategy shown in Eq. (\ref{eq:strategy}), here we replace the integration over $\bm{k}$ with an integration over energy, for example:
\begin{equation}
	\begin{split}
		& \int\frac{\rd^2k}{(2\pi)^2}\frac{f(\bm{k})}{\omega+i/(2\tau)-\epsilon_+(\bm{k},z)}\\
		\approx ~& \nu \int \rd \zeta \lrangle{\frac{f(\bm{k})}{\omega+i/(2\tau)-\zeta}}_{\text{FS}_+}, 
	\end{split}
\end{equation}
where the average is over the Fermi surface determined by $\epsilon_+(\bm{k},z)=0$. Following this strategy, in the leading order in $t^2$, the $c$-axis conductivity is obtained as
\begin{equation}
\label{eq:Drude}
		\sigma_c(\theta)=\lrangle{\frac{ 4\nu q^2_et^2\tau }{1+\tau^2\mu^2 f^2(\theta)}+\frac{ 4\nu q^2_et^2\tau }{1+\tau^2\mu^2 f^2(\pi/2-\theta)}}_{\text{FS}},
\end{equation}
where the function $f(\theta)$ takes the expression
\begin{equation}
\label{eq:Drude1}
	\begin{split}
		& f(\theta)=\left(\sqrt{\xi\gamma}\sin\theta+\gamma\cos\varphi\right)^2\\
		& \quad \quad \quad +\left(-\sqrt{\frac{\xi}{\gamma}}\cos\theta+\frac{1}{\gamma}\sin\varphi\right)^2-1, \\
		& \gamma = \sqrt{\frac{m_y}{m_x}}, \quad \epsilon_0 = \frac{\left(q_e B a_0\right)^2}{\sqrt{m_xm_y}}, \quad  \xi=\frac{\epsilon_0}{2\mu},
	\end{split}
\end{equation}
and $\lrangle{\cdots}_{\text{FS}}$ is simply the average over $\varphi\in[0,2\pi]$. A schematic plot of Eq. (\ref{eq:Drude}) and (\ref{eq:Drude1}) in terms of the resistivity $\rho^n_c=1/\sigma_c$ is shown in Fig. \ref{fig:resisqu}. As can be seen from Fig. \ref{fig:resisqu}, although it presents the desired period and locations of maxima/minima, it has a qualitatively different shape near the maxima and minima. To make it even worse, in order to produce the amplitude of oscillation in $\rho^n_c$ observed in experiments, we have to adopt an effective electron mass on the order of $10^{-5}$ the bare electron mass, which is unrealistic. In this sense, the experimental data favor the Dirac fermions over the conventional fermions.

\section{Materials and Methods}
\label{sec:MM}

Here we discuss the materials and methods for obtaining the experimental data. Single crystals of LBCO with $x=1/8$ studied here were grown in an infrared image furnace by the floating-zone technique. They are pieces from the same cylindrical crystal used previously to characterize two-dimensional fluctuating superconductivity \cite{PhysRevLett.99.067001}. Single crystal samples were cut and aligned into slabs, then fixed on a 0.5 mm thick sapphire substrate. The imperfection in the sample alignment, estimated from X-ray diffraction, is less than 0.5 degrees. For transport measurements, current contacts were made at the ends of the longest dimension of crystals to ensure uniform current flow, while the voltage contacts were made on both the top and side of the crystals. We used a low-temperature contact annealing procedure \cite{PhysRevLett.99.067001} leading to low contact resistance $(<0.2 ~\Omega)$ that allows us to measure the resistivity over seven orders of magnitude. The angular-resolved magnetoresistance (ARMR) was measured using the 4-point probe in-line method in a Quantum Design Physical Property Measurement System (PPMS) equipped with a 14 T superconducting magnet. The resistivity measurements have been performed with the current applied along either the in-plane direction or the c-direction using DC and AC transport options with a current range of 50 $\mu \mathrm{A}-1 \mathrm{~mA}$. Both DC and AC methods produced the same results. The data shown are from the AC transport measurements $(17 \mathrm{~Hz})$. For crystal alignment with the magnetic field, horizontal and vertical sample rotators with an angular resolution of $\sim 0.1^{\circ}$ were used. Temperature-dependent ARMR data were taken from $1.8$ K to $300 \mathrm{~K}$, at various fields up to $14 \mathrm{~T}$. ARMR data at fixed temperatures and magnetic fields were taken in-situ with a vertical sample rotator as a function of the in-plane magnetic field angles $(\theta)$ in a range of $-15^{\circ}$ to $355^{\circ}$.

\bibliography{Conpaper}

\end{document}